\documentclass[11pt]{article}

\usepackage{acl}

\usepackage{times}
\usepackage{latexsym}
\usepackage{amsmath}
\usepackage[T1]{fontenc}

\usepackage[utf8]{inputenc}

\usepackage{microtype}

\usepackage{inconsolata}

\usepackage{graphicx}
\usepackage{booktabs}
\usepackage{amssymb}
\usepackage{multirow}
\usepackage{tcolorbox}
\tcbuselibrary{breakable,skins}
\usepackage{algorithm}
\usepackage{algpseudocode}

\algrenewcommand\algorithmiccomment[1]{\hfill$\triangleright$~\footnotesize#1}

\title{Listen, Pause, and Reason: Toward Perception-Grounded Hybrid Reasoning for Audio Understanding}

\author{
 \textbf{Jieyi Wang\textsuperscript{1,2}},
 \textbf{Yazhe Niu\textsuperscript{1,3}}\thanks{Corresponding author.},
 \textbf{Dexuan Xu\textsuperscript{2}},
 \textbf{Zhongyu Wei\textsuperscript{4}}
\\
 \textsuperscript{1}Shanghai AI Laboratory,
 \textsuperscript{2}Peking University,
 \textsuperscript{3}CUHK MMLab,
 \textsuperscript{4}Fudan University
\\
 \small{
   {joysw@stu.pku.edu.cn, niuyazhe314@outlook.com}
 }
}

\begin{document}
\maketitle
\begin{abstract}
Recent Large Audio Language Models have demonstrated impressive capabilities in audio understanding.
However, they often suffer from perceptual errors, while reliable audio reasoning is unattainable without first grounding the model’s perception in structured auditory scenes.
Inspired by Auditory Scene Analysis, we first introduce a Perception-Aware Question Answering (PAQA) dataset.
PAQA implements a hierarchical decoupling strategy that separates speech from environmental sound and distinguishes multiple speakers, providing explicit perceptual reasoning for training.
Building on this, we propose HyPeR, a two-stage Hybrid Perception-Reasoning framework.
In Stage I, we finetune the model on PAQA to perceive acoustic attributes in complex audio.
In Stage II, we leverage GRPO to refine the model's internal deliberation.
We also introduce PAUSE tokens to facilitate latent computation during acoustically ambiguous phases and design perceptual consistency reward to align reasoning rationales with raw audio.
Experiments across benchmarks demonstrate that HyPeR achieves absolute improvements over the base model, with performance comparable to large-scale models, stressing the effectiveness of hybrid perception-grounded reasoning for robust and multi-speaker audio understanding. Our code and data is available at \url{https://github.com/JOY-SWang/HyPeR}.


\end{abstract}

\section{Introduction}
Recent Large Audio Language Models (LALMs) have made strides in audio understanding~\citep{Qwen2Audio,AudioFlamingo,SALMONN,xue2025hhcodec}, with steady progress on challenging audio reasoning benchmarks~\cite{MMAU, MMAR}.
Yet, their performance is dominantly capped by perceptual errors, where the models struggle with distinguishing environmental sounds, and accurately transcribing speech.
Although LALMs have further made notable progress in reasoning via Chain of Thought (CoT)~\citep{AudioReasoner,AudioCoT} and reinforcement-learning (RL) post-training~\citep{R1AQA,AudioThinker}, the reasoning paths produced upon unreliable perceptions may hallucinate evidence and bring about bad comprehension in Audio Question-Answering (QA)~\citep{pass1yy}. Moreover, current models often derive answers primarily from text-based reasoning without acoustic evidence, leading to weak audio grounding.

\begin{figure}[t]
\centering
\includegraphics[width=1\linewidth]{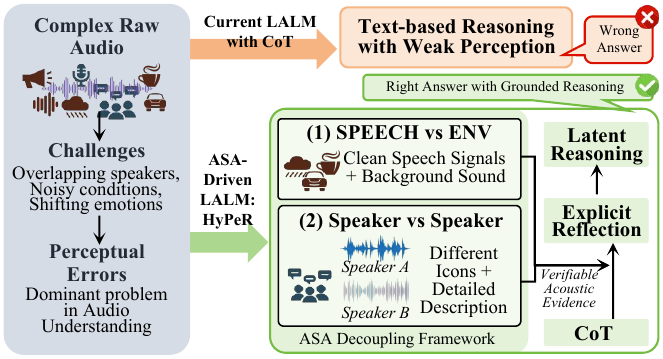}
\caption{ASA-inspired layered decoupling for perception-grounded audio reasoning. Rather than directly mapping audio to text, we separate background sound from speech and distinguish multiple speakers to construct verifiable acoustic evidence, and then perform grounded reasoning on top of this evidence.}
\label{fig:mov}
\end{figure}

Previous research on audio grounding centered on Sound Event Detection with on- and off-set timestamps~\citep{xu2021text} and interval localization~\citep{ghosh2024gama, TwS}, which brings about additional architectural complexity and extra inference time. Furthermore, it's hard for current LALMs to follow the routine since they may exhibit temporal misalignment~\cite{kuan2025can}. To address these limitations, we focus on verifiable acoustic attributes and source-aware cues to improve audio grounding. Drawing inspiration from \textbf{Auditory Scene Analysis (ASA)}, the human brain processes complex soundscapes through layered decoupling pathways~\cite{ASAbregman1994auditory, michelsanti2021overview}, effectively segregating the background sound (ENV) from the foreground one (SPEECH) and distinguishing multiple speakers before performing high-level semantic synthesis, as shown in Figure~\ref {fig:mov}. 

However, directly applying LALMs to background sound recognition remains unsatisfactory in practice. Specialized audio–text alignment models (e.g., CLAP~\cite{elizalde2023clap, elizalde2024natural,ghosh2025reclap,niizumi2024m2d}) report mean Average Precision (mAP) values below 50\% on FSD50K, a multi-label audio tagging dataset, while Qwen2-Audio-7B-Instruct only achieves 15\% mAP in our experiment. To address this gap, we introduce \textbf{PAQA}, a dataset specifically designed to benchmark and facilitate this decoupling. PAQA focuses on two core disambiguations: (1) \textbf{Speech vs. Environment}: isolating linguistic signals from non-speech interference; and (2) \textbf{Speaker vs. Speaker}: resolving multi-party attribution to recover conversational dynamics. PAQA contains 7,470 multiple-choice Audio-QA pairs, each enriched with structured annotations, including background-music separation, speaker analysis, and multi-turn reflections. By recording both internal acoustic cues and final responses, PAQA forces the model to ground its reasoning in explicit perceptual evidence.

To better detect and ground perceptual cues and acoustic attributes, we propose \textbf{HyPeR}, a two-stage \textbf{Hy}brid \textbf{P}erception-\textbf{R}easoning framework that unifies explicit reflective reasoning with implicit latent computation. Explicit Perception in Stage I involves Supervised Fine-Tuning (SFT) on PAQA to teach the model to imitate human-like layered auditory decomposition. Nevertheless, we observe that the generated CoT often remains imprecise when describing certain acoustic attributes (e.g., tone, pitch, background noise texture, and paralinguistic emotion). Inspired by~\citet{goyal2023think}, We mimic the ``think before speak'' pattern and introduce a special token, \texttt{<PAUSE>}, to represent a latent reasoning step during inference in which no output token is extracted or autoregressively fed back, enabling the model to perform additional reasoning via Group Relative Policy Optimization (GRPO) before committing to verbal descriptions of difficult acoustic attributes. Moreover, we empirically find that when the model is about to generate tokens related to the acoustic keyword set, the token selection confidence is often lower. To better place the <PAUSE> token, we propose a sliding-window group confidence~\cite{wang2025deepthink} to detect locally unreliable spans during generation. The reward function is designed for audio grounding and jointly balances answer correctness, reasoning consistency, and format compliance. Our experimental results on PAQA and other benchmarks demonstrate that HyPeR significantly reduces perceptual errors and achieves strong performance on complex audio understanding and reasoning tasks, particularly in noisy speech and multi-speaker scenarios.

Our contributions are summarized as follows:
\begin{itemize}

\setlength{\leftmargin}{0pt}
\vspace{-8pt}
\item We focus on the Perception-Grounded Audio Understanding and redefine the reasoning of LALMs from a direct audio-to-text mapping to CoT with explicit acoustic grounding on environment sound and multi speakers based on Auditory Scene Analysis.
\vspace{-8pt}
\item We introduce PAQA, a novel benchmark designed to operationalize this hierarchical reasoning, with stepwise reasoning and reflection annotations across multi-speaker QA, noisy speech translation, and environment-centric QA, intended to suppress shortcut learning and promote acoustic grounding.
\vspace{-8pt}
\item We propose HyPeR, a hybrid framework that unifies explicit reflection with latent reasoning, with PAUSE token detecting acoustic attributes. By employing a GRPO-based reinforcement learning strategy with multi-dimensional rewards (accuracy, consistency, and grounding), HyPeR effectively bridges the perception-reasoning gap.
\end{itemize}

\section{Related Works}
\subsection{Large Audio–Language Models (LALMs)}
Early LALMs such as Qwen2-Audio-7B-Instruct~\citep{Qwen2Audio}, Audio Flamingo~\citep{AudioFlamingo}, and SALMONN~\citep{SALMONN} advanced ASR, but remained fragile in real-world reasoning tasks involving multi speakers and non-stationary noise. More recent omni-/speech-native systems broaden the interface beyond transcripts with end-to-end audio generation such as OpenAI’s GPT-4o Audio models~\citep{GPT4oAudioPreview}, and Gemini 2.5 Pro~\citep{Gemini25}. However, on-demand CoT in Audio Flamingo~3\citep{Chen2025AudioFlamingo3} and structured CoT in Audio-Reasoner~\citep{AudioReasoner}, yet models often reverted to transcript shortcuts whenever acoustic evidence was difficult to verbalize. Recent work~\citep{ghosh2024gama, TwS} has therefore shifted toward architectural audio evidence alignment and multi-representation fusion, but brings about additional architectural complexity and extra inference time. To address these limitations, we release a structured dataset that couples multi-speaker and background-rich audio, explicitly guiding LALMs to ground decisions in acoustic rather than text.

\subsection{Explicit Reasoning in LLMs}

In LLMs, structured reasoning through CoT, reflection, and RL post-training has yielded consistent gains beyond SFT~\citep{Guo2025DeepSeekR1}. While Vision-R1~\citep{Huang2025VisionR1} and Video-R1~\citep{Feng2025VideoR1} extended RL-based reasoning to overthinking suppression. In audio, GRPO-style RL underlies R1-AQA and Omni-R1~\citep{Shao2024DeepSeekMathGRPO,R1AQA,OmniR1}, with mixed evidence on whether RL alone suffices. More recent approaches~\citep{SARI,AudioThinker,R1AQA,SearchR1} highlight that objectives should reward useful and concise reasoning rather than verbosity. In this work, we instead unify explicit, audio-grounded reasoning with reflection, operationalized through a multi-term reward that enforces correctness and conciseness.

\subsection{Implicit Latent Reasoning and PAUSE-Triggered Computation}

Complementary to explicit rationales, implicit computation allocates additional internal processing before token emission. Learned \texttt{<PAUSE>} tokens can trigger silent forward passes ~\citep{goyal2023think}, echoing earlier adaptive-computation approaches~\citep{ACT,PonderNet} that learn instance-dependent halting policies. To our knowledge, such latent computation has not been systematically validated in audio–language reasoning. Our contribution is to extend \texttt{<PAUSE>} to LALMs and couple it with a lowest-group-confidence (LGC) controller: when confidence drops on acoustically inexpressible cues, HyPeR diverts into a short, budgeted latent stream and can abort tail trajectories under severe uncertainty.

\section{Data with Audio Layered Decoupling}

\subsection{ASA-Inspired Taxonomy}
To bridge the gap between raw acoustic signals and high-level complex reasoning, we introduce the PAQA dataset,  which is designed to supervise the perception-reasoning decoupling process itself, providing explicit "Perceptual Traces" based on Auditory Scene Analysis~\cite{ASAbregman1994auditory}. We further analyze Qwen2-Audio-7B-Instruct’s bad output cases on the CoTA~\citep{AudioReasoner} benchmark and identify two major challenges. 

\paragraph{Level 1: Speech vs. Environment (S-E)}
To prevent the model from misattributing background interference as conversational evidence, we synthesize complex auditory scenes using MUSAN~\citep{snyder2015musan} and FSD50K~\cite{fonseca2021fsd50k}. For a speech clip $s$ and an environmental noise $n$, we apply RMS-normalization and mix them with a dynamic $SNR$ range of [0,20] dB. Crucially, each item is annotated with an Environment Tag (e.g., \textit{"Background: Rain and distant traffic"}), forcing the model to distinguishing speech and non-speech during the reasoning phase.

\paragraph{Level 2: Speaker vs. Speaker (S-S)}
To resolve multi-party conversational structures, we annotate speaker turns using a structured format. To ensure the model performs true Speaker Attribution rather than shortcutting via global transcripts, we introduce the Quote-Presence Test (QPT).
QPT measures the alignment between the model's attributed speaker segments and the raw ASR output (checked by Qwen3-ASR~\citep{shi2026qwen3}). We filter out items with QPT<0.85 to ensure the reasoning is strictly grounded in the temporal sequence of the audio. The alignment is formulated as: 

\begin{equation}
    \mathrm{QPT} = \frac{1}{M} \sum_{i=1}^{M} \max_{1 \le j \le N} \phi(\hat{s}_i, \hat{a}_j),
\end{equation}
where $\hat{s}$ and $\hat{a}$ denote the normalized strings of attributed sentences and ASR snippets, respectively. $\Phi(\bullet)$ computes the fuzzy overlap ratio (SeqRatio) between two strings. Detailed data sources, audio mixing procedure, SNR setting, and multi-speaker synthesis pipeline are provided in Appendix \ref{sec:dataAna}.
  
\begin{figure*}[htp]
\centering
\includegraphics[width=0.9\linewidth]{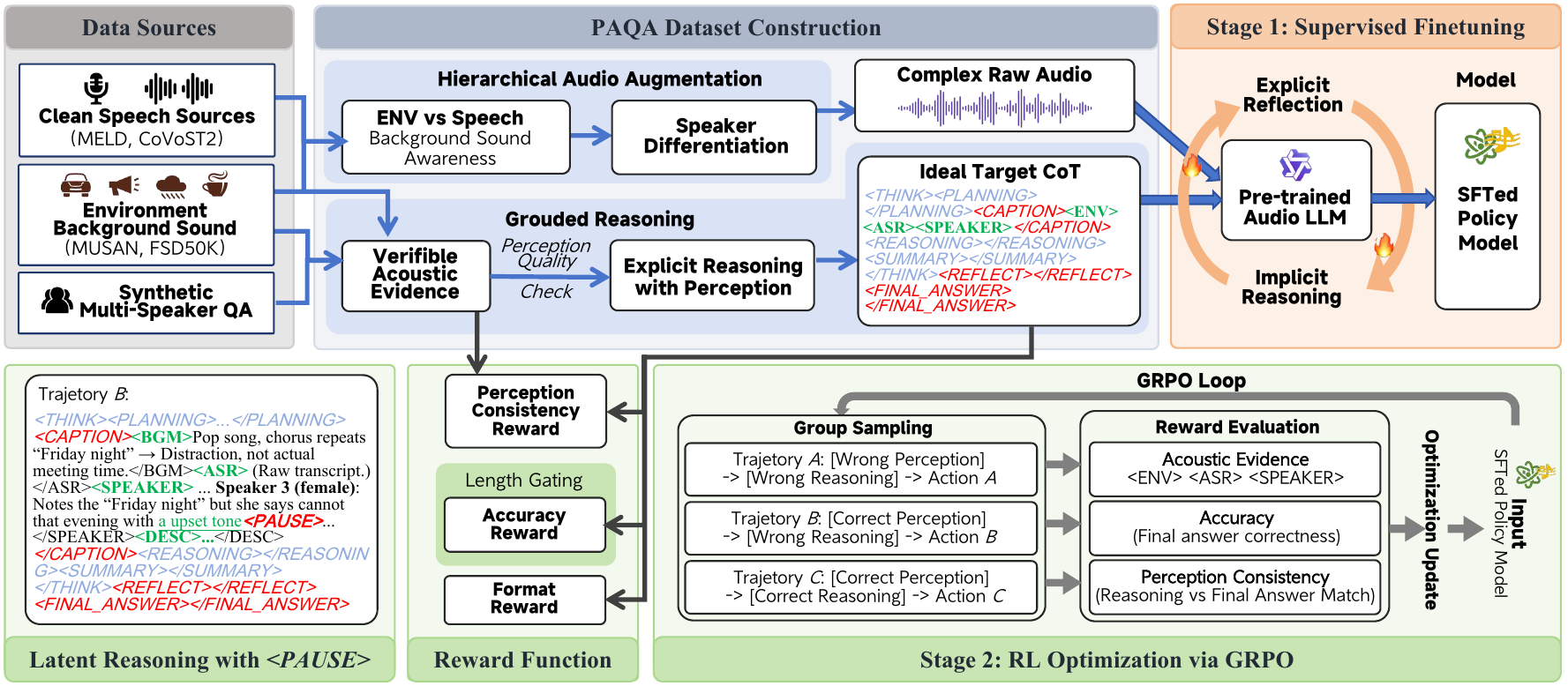}
\caption{An overview of our framework HyPeR. First, we construct the PAQA dataset with complex audio collection and hierarchical audio augmentation. Then each training example is converted into a grounded reasoning Chain-of-Thought containing verifiable acoustic evidence for explicit reasoning.
In Stage 1, a pretrained audio language model is optimized on PAQA for structured reasoning that explicitly links perception to reasoning. To handle acoustic cues, HyPeR further introduces implicit reasoning with <PAUSE>, which allows HyPeR to listen, pause, and then reason.
In Stage 2, the SFT-initialized policy model is further improved through group-based rollout sampling and multi-objective reward optimization.
The example shown that HyPeR can mark repeated background lyrics as low-confidence distraction, reflect on the mistake, and then revise the answer based on grounded perception.  
}
\label{fig:pipeline}
\vspace{-6pt}
\end{figure*}

\subsection{Data Collection \& Statistics}
\label{sec:dataR}
In natural conversation, speakers frequently self-monitor and revise their utterances. Building on prior work showing that reflection-driven self-correction improves model performance in reasoning tasks~\citep{shinn2023reflexion,madaan2023selfrefine,wang2023selfconsistency}, we adopt a reflection-augmented pipeline for complex audio understanding. Concretely, a lightweight baseline model first generates an initial \texttt{<RESPONSE>} for each audio QA item, as illustrated in Figure ~\ref{fig:pipeline}. We then automatically detect errors, such as option mismatches, speaker attribution mistakes, hallucinated content inconsistent with ASR transcripts, or misinterpretation of noise cues. Finally, we prompt the model to produce a grounded diagnostic analysis \texttt{<REFLECT>} with manual check. This analysis explicitly references \texttt{<BGM>}, \texttt{<SPEAKER>}, and \texttt{<ASR>} to explain the failure and localize the supporting evidence. Conditioned on this analysis, the model is guided to generate a corrected \texttt{<FINAL\_ANSWER>}. For training, we store the triplet (\texttt{<RESPONSE>}, \texttt{<REFLECT>}, \texttt{<FINAL\_ANSWER>}), which provides explicit reflection supervision and, from each original audio item, yields an additional corrected example, effectively doubling the supervised data while enriching them with interpretable, perception, grounded self-correction signals. A full statistical breakdown of PAQA is given in Appendix \ref{app:dataAnalysis}. The prompt template used to construct reflection-augmented supervision is provided in Appendix \ref{sec:prompt}.  
\section{Method}
\subsection{Overall Architecture}
To bridge the gap between low-level acoustic perception and high-level audio-linguistic reasoning, we propose \textbf{HyPeR}, a unified Hybrid Perception-Reasoning framework that mimics the human brain’s hierarchical processing of auditory scenes. Given an audio input $X_a$ and a textual query $Q$, HyPeR aims to generate a logically grounded response $Y$. We decompose this into a two-stage hierarchical process: Explicit Perceptual Reflection and RL-driven Latent Reasoning.

We first enhance the model's perception through SFT on PAQA dataset. 
The model is trained based on Qwen2-Audio-7B-Instruct to generate a reasoning chain that explicitly performs decoupling: first identifying the acoustic environment (Speech vs. Environment) and then resolving speaker dynamics (Speaker vs. Speaker).
These traces, encapsulated within \texttt{<REFLECT>} tags, serve as the "logical grounding" for the final answer.
Besides, recognizing that non-textualizable acoustic nuances (e.g., subtle prosodic shifts or overlapping textures) are difficult to describe explicitly, we introduce the \texttt{<PAUSE>} token, which represents an inference step in which no visible token is produced and no token is fed back autoregressively, allowing the model to carry out latent reasoning internally.
During the RL stage, the model learns to invoke this token autonomously when its confidence is low.
This allows dynamic latent reasoning, where the model allocates additional internal computation to refine its latent states before generating traces or the final response.

\subsection{Stage I: Explicit Perception (SFT)}
In this stage, the model is trained via Supervised Fine-Tuning (SFT) on the PAQA dataset to imitate human-like auditory decomposition. Following a structured reasoning pipeline, the model generates an explicit trace $T$ consisting of four sequential components: (1) Planning (P): Outlining the logic required to address the query. (2) Captioning (C): Extracting multi-modal information, especially multi-layered acoustic features, including environment (\texttt{<ENV>}), speaker dynamics (\texttt{<SPEAKER>}), and speech content (\texttt{<ASR>}). (3) Reasoning (R): Performing step-by-step analytical deduction based on P and C. (4) Summary (S): Synthesizing the reasoning into a concise internal conclusion. (5) Reflection (R'): Producing a transparent analysis of background sound and speaker, and reflection that allows for direct inspection of the summary to a better answer. This process is formalized in Eq.\ref{eq:seq}.

\begin{equation}
\begin{aligned}
P &\sim f_\theta(\mathbf{X}_a, \mathbf{Q}), \\
C &\sim f_\theta(\mathbf{X}_a, \mathbf{Q}, P), \\
R &\sim f_\theta(\mathbf{X}_a, \mathbf{Q}, P, C), \\
S &\sim f_\theta(\mathbf{X}_a, \mathbf{Q}, P, C, R), \\
R' &\sim f_\theta(\mathbf{X}_a, \mathbf{Q}, P, C, S).
\end{aligned}
\label{eq:seq}
\end{equation}

The explicit trace $T=\{P,C,R,S, R'\}$ serves as the logical perceptual grounding for the final answer. We aim to teach the model to generate its responses in a specific, structured format, it lays the groundwork for the subsequent reinforcement learning phase. The optimization goal of this stage is the standard cross entropy loss in Equation~\ref{eq:sft}.

\begin{equation}
    \mathcal{L}_{\text{SFT}} = -\sum_{i=1}^{|\mathbf{T}|} \log P(t_i \mid \mathbf{X}_a, \mathbf{Q}, \mathbf{T}_{<i})
\label{eq:sft}
\end{equation}

\subsection{Confidence-based Transition Gating}
After generating the explicit trace $T$, HyPeR evaluates whether the acoustic information has been sufficiently resolved. Audio streams contain a host of non-verbal cues, such as speaker intonation, overlapping speech, and ambient noise, that are often difficult to fully articulate in explicit text. We found a connection between the reasoning trace's lower confidence score and non-verbal cues. 
Therefore, we consider the Lowest Group Confidence (LGC) metric $C_t$ at each decoding step $t$. Each token $t$ is linked to a sliding window group $K_i$, consisting of $n$ previous tokens. In particular, we identify its bottom 15\% group confidence. For each window, we compute a normalized  mean probability:

\begin{equation}
    C_{K_i}=\frac{1}{|K_i|}\sum_{t \in K_i}C_t,
\end{equation}
where $|K_i|$ is the number of tokens in group $K_i$. The LGC of the trajectory is then defined as the minimum of these window confidence scores, $\mathrm{LGC}(\mathbf{y}) = \min_{k=1,\dots,K} C_{K_i}$. 
This definition emphasizes the weakest local segment within the reasoning trajectory: even a small cluster of highly uncertain tokens can significantly reduce LGC, making it a sensitive indicator of detecting local reasoning collapse, a phenomenon effectively demonstrated by \citet{wang2025deepthink}. 

When the LGC falls into the intermediate ambiguity range $(\tau_{abort}, \tau_{PAUSE}]$, the model triggers a "Think-Before-Speak" reasoning step. If LGC drops below $\tau_{abort}$, the model autonomously aborts the trajectory to prevent unproductive reasoning loops or hallucinations, significantly accelerating inference by pruning unpromising paths.

\subsection{Latent Reasoning with PAUSE Token}
During the initial phase of Stage II training, we introduce a keyword-based heuristic to calibrate the model's sensitivity to acoustic nuances. We maintain a keyword set $K$={"tone", "pitch", "noise", "emotion", …} representing non-textualizable cues. Whenever a word $w \in T$ appears in the recent context, we apply a positive logit bias $\beta_{ac}>0$ to the <PAUSE> token, as shown in Figure \ref{fig:framwork}:
\begin{equation} 
\ell_{\texttt{<PAUSE>}} \leftarrow \ell_{\texttt{<PAUSE>}} + \beta_{\text{ac}} \cdot \mathbb{I} \big[ \exists w \in \mathcal{K} \big] 
\end{equation}

This mechanism serves as a cold-start prior for the threshold $\tau_{abort}$, encouraging the model to allocate latent computation specifically when the explicit text involves speech-only cues. 


When a PAUSE is triggered at step $t$, the model emits a \texttt{<PAUSE>} special token and generates a sequence of latent tokens $\hat{\mathbf{z}}_{1:L}$. Crucially, these tokens function as a non-volatile computational cache; they are not surfaced in the final visible output and are explicitly excluded from the gradient calculations during the generation of the final response to maintain efficiency. Their function is only to iteratively update and refine the model’s internal hidden state $H_t$, enabling a deeper, more grounded processing of complex audio features before resuming the generation of visible tokens. The relationship between the full internal sequence $\tilde{\mathbf{y}}$ and the visible output $y_{vis}$ is formalized as: 

\begin{equation} 
\tilde{\mathbf{y}} = \mathbf{y}_{1:t^\star} \oplus \texttt{<PAUSE>} \oplus \hat{\mathbf{z}}_{1:L}, ~ \mathbf{y}_{\mathrm{vis}} = \mathbf{y}_{1:t^\star} 
\end{equation}

The architecture ensures the model "thinks" internally as it processes intricate auditory scenes, effectively bridging the gap between low-level acoustic perception and high-level text reasoning.

\subsection{Stage II: GRPO-based RL Training}
While SFT in Stage I establishes a structural foundation for auditory decomposition, its efficacy is inherently limited by the nature of imitation learning. To optimize the model's internal reasoning ability, we introduce a second stage of optimization using Group Relative Policy Optimization (GRPO) \citep{Shao2024DeepSeekMathGRPO} from the SFT checkpoint as the reference policy $\pi_{\text{ref}}$ frozen.
We generate groupwise rollouts, compute $R(\mathbf{z})$ via ~\eqref{eq:reward}, and update $\pi_{\theta}$ with GRPO~\citep{Shao2024DeepSeekMathGRPO}. We partition rollouts by task group $g\in\{\text{PAQA},\text{AVQA}\}$. For each trajectory $i$ within a group, we compute the relative advantage to reduce variance: 

\begin{equation} \tilde{R}^{(i)} = R^{(i)} - \frac{1}{m_g}\sum_{j\in g} R^{(j)},
\end{equation} 
where $m_g$ is the number of samples in the group.

To specifically address the "thinking" process regarding non-textual audio cues, we utilize the keyword set $K$ (e.g., "tone", "pitch", "noise") as a cold-start prior. In early RL iterations, these keywords provide initial guidance on acoustic sensitivity by influencing the gating threshold $\tau_{PAUSE}$. Crucially, we incorporate the Lowest Group Confidence (LGC) metric $C_t$ into the advantage calculation. The LGC serves as a proxy for the "logical weakest link" in a reasoning trajectory. For a trajectory $i$ with a raw task reward $r^{task}_i$ (encompassing accuracy, formatting, and consistency), the weighted advantage $A_i$ is defined as: 
\begin{equation} 
A_i = w_i \cdot (r_i^{\text{task}} - \bar{r}),
\end{equation} 
where $w_i=clip(std(LGC(y)))$ is a standardized weight derived from the trajectory's LGC. Here, $w_i=0$ for trajectories that fall below the $\tau_{abort}$ threshold, effectively pruning unpromising or unstable reasoning paths during optimization.

\subsection{Multi-Objective Reward Function}
To ensure the model not only produces accurate answers but also generates interpretable, perception-grounded reasoning, we design a composite reward function $R$. It is defined as a weighted sum of four specialized components:

\begin{equation}
\begin{aligned}
    R= & w_{\text{acc}} \, \mathcal{R}_{\text{acc}} \;+ \; w_{\text{cons}} \, \mathcal{R}_{\text{cons}}(\hat{y}, \hat{y}_{\text{CoT}})\;+\; \\
&  w_{\text{fmt}} \, \mathcal{R}_{\text{fmt}} \;+\;  w_{\text{len}} \, (\mathcal{R}_{\text{acc}} \times \mathcal{R}_{\text{len}}),
\end{aligned}
\label{eq:reward}
\end{equation}

where $\mathcal{R}_{\text{acc}}$ and $\mathcal{R}_{\text{fmt}} $ provide the fundamental supervision for task completion, while $\mathcal{R}_{\text{cons}}(\hat{y}, \hat{y}_{\text{CoT}})$  and $\mathcal{R}_{\text{len}}$ serve as perceptual and structural regularizers to stabilize the learning of the hybrid reasoning process.

\subsubsection{Accuracy and Format Rewards}
The Accuracy Reward ($\mathcal{R}_{\text{acc}}$) is a binary signal $\mathbf{1}[\hat{y} = y]$. We prioritize extracting $\hat{y}$ from the \texttt{<FINAL\_ANSWER>} tag, with a fallback to the \texttt{<RESPONSE>} tag to ensure robustness during early RL stages. 
The Format Reward ($\mathcal{R}_{\text{fmt}}$) addresses the reward sparsity inherent in complex structural requirements. To prevent "gradient collapse" where the model fails to receive any signal due to strict schema violations and other hacking cases, we adopt a progressive format shaping strategy. We reward a "weak format" (correct \texttt{<THINK>} and \texttt{<RESPONSE>} sequence) with a base score, while the "strict format" (inclusion of specific environment and speaker tags) is implicitly incentivized through the consistency rewards described below.
 
\subsubsection{Perceptual Consistency Reward}
To enforce the "perception-grounded" nature of our framework, $\mathcal{R}_{\text{con}}$ regularizes the reasoning chain along three acoustic-logical axes:

\paragraph{BGS Robustness.} To eliminate illusions where the model treats background sound as causal evidence for speech-related questions, we define a background sound gate $\mathcal{r}_{\text{bgs}}$. If the reasoning chain invokes environmental cues (e.g., "the background music suggests...") as a causal basis for linguistic content, $\mathcal{r}_{\text{bgs}}$ is set to 0; otherwise, it is 1.

\paragraph{Speaker-ASR Fidelity.} Within the \texttt{<THINK>} block, we extract speaker-attributed quotes $S={s_i}$ and verify them against the raw ASR transcript $A={a_j}$. We define the fidelity score $\mathcal{r}_{\text{fid}}$ as: 

\begin{equation} r_{\text{fid}} = \frac{1}{|\mathcal{S}|} \sum_{s \in \mathcal{S}} \max_{a \in \mathcal{A}} \phi(\hat{s}, \hat{a}), 
\end{equation} 
where $\phi$ is the character-level Levenshtein similarity. This ensures that the model's "perception" is strictly anchored to the acoustic evidence rather than hallucinated text.

\paragraph{Reasoning-Answer Alignment.} We reward the agreement between the model's internal conclusion $\tilde{y}$ in reasoning CoT and its final text answer $\hat{y}$. The final consistency reward can be defined as: 

\begin{equation} \mathcal{R}_{\text{cons}} = \mathcal{r}_{\text{bgs}} \cdot \left( \lambda_{\text{fid}} r_{\text{fid}} + \lambda_{\text{align}} r_{\text{align}} \right). \end{equation}

\subsubsection{Length Shaping via Correctness Gating}
\label{sec:lengthReward}
To prevent "reasoning collapse" (too short) or "superficial verbosity" (too long), we introduce $\mathcal{R}_{\text{len}}$, which is only activated when $\mathcal{R}_{\text{acc}}=1$. We use a piecewise-linear function with a penalty for completions exceeding $T_{max}$ tokens or failing to reach $T_{min}$ tokens. Crucially, any content generated after the \texttt{</FINAL\_ANSWER>} tag results in a zeroed length reward to encourage clean termination~\cite{arora2025training}.

\section{Experiments}
\subsection{Implementation Details}
All experiments fine-tune the same pretrained base model (Qwen2-Audio-7B-Instruct), using the framework introduced by~\citet{R1AQA} and ~\citet{lightrft} separately for SFT and RL training.
Training is conducted with a batch size of 1 per GPU, with by 2 gradient accumulation steps, resulting in an effective total batch size of 16.
We adopt a learning rate of $1e-6$, a temperature of 1.0, and configure the GRPO to sample 8 responses per group with a KL coefficient $\beta$ of 0.1. 
For models incorporating PAUSE a latent thinking mechanism, we set $\tau_{PAUSE}=0.5$ and allow up to 3 PAUSEs per sequence with 64 thinking tokens each, plus $\tau_{abort}=0.05$ for think token containment.

\subsection{Benchmarks and Metrics}
We evaluate six configurations: \textbf{SFT}, standard fine-tuning; \textbf{GRPO-Nothink}, GRPO post-training without \texttt{<REFLECT>} or \texttt{<PAUSE>}; \textbf{GRPO+CoT}, GRPO enhanced with thinking before the answer (in the weak format of \texttt{<THINK><ANSWER>}); \textbf{GRPO+ExpCoT}, GRPO enhanced with explicit \texttt{<THINK>} (including \texttt{<REFLECT>}) but no \texttt{<PAUSE>}; \textbf{Ours (HyPeR)}, GRPO enhanced with the explicit schema and \texttt{<PAUSE>}; and \textbf{External Baselines} including GPT-4o Audio~\cite{jaech2024openai}, Gemini 2.5 Flash~\cite{comanici2025gemini}, Audio-Flamingo-3~\cite{Chen2025AudioFlamingo3}, OmniVinci~\cite{omnivinci2025}, Qwen2.5-Omni~\cite{xu2025qwen2}, and existing LALM reasoning frameworks like Audio-Reasoner~\citep{AudioReasoner}, Audio-CoT~\citep{AudioCoT} and Audio-Thinker~\citep{AudioThinker} (all trained on Qwen2-Audio-7B-Instruct). 

We use PAQA (train set) for supervised finetuning. For RL training, we utilize 30,000 augmented samples generated upon the AQVA~\citep{yang2022avqa} dataset, with each response reformulated into a \texttt{<think>...</think><answer>...</answer>} reasoning–answer structure.
Models are evaluated on several benchmarks, \textbf{PAQA Test} (``MSQA-hard" for the subset of QA with >3 speakers, ``ENVQA-hard" for the subset with background sound under SNR=5dB), \textbf{MMAU}~\citep{MMAU}, \textbf{MMAR}~\citep{MMAR}, and \textbf{MMSU}~\citep{wang2025mmsu}. The results are listed below and in the Appendix.\ref{app:moreExp}.

\subsection{Direct LALM Percepting Underperforms} 
To evaluate LALM’s perception ability, we first use models directly recognizing background sound on FSD50K dataset, a multi-label sound event classification benchmark, and calculate Word Error Rate (WER) and Character Error Rate (CER) based on the transcripts generated in the explicit reasoning on the PAQA test dataset. Qwen2-Audio-7B-Instruct achieves only 14.7\% mAP on FSD50K, far below the audio–text alignment model CLAP23\citep{elizalde2023clap} 's 50\%, and poor for direct generation in multi-label environmental sound tagging. HyPeR narrows the gap to 43.6\% and achieves a remarkably low WER of 1.65\% and CER of 1.61\%, demonstrating that our fine-tuned model’s reasoning is grounded in more accurate perception, ruling out hallucination.

\begin{table}[t]
\centering
\small
\caption{Diagnostic evaluation of perception ability.
FSD50K evaluates sound event classification using mAP (\%).
PAQA evaluates transcript quality in explicit reasoning using WER/CER (\%, lower is better).}
\label{tab:perception_diagnostic}
\begin{tabular}{lccc}
\toprule
& \multicolumn{1}{c}{\textbf{FSD50K}} & \multicolumn{2}{c}{\textbf{PAQA}} \\
\cmidrule(lr){2-2} \cmidrule(lr){3-4}
\textbf{Model} & \textbf{mAP $\uparrow$} & \textbf{WER $\downarrow$} & \textbf{CER $\downarrow$} \\
\midrule
HyPeR (Ours)        & 43.6         & \textbf{0.78} &  \textbf{0.62}   \\
base model  & 14.7          &  0.87  &    0.78       \\
CLAP23              & \textbf{48.6} &  23.07   &    24.80            \\
\bottomrule
\end{tabular}
\end{table}

\begin{figure*}[ht]
\centering
\includegraphics[width=1\linewidth]{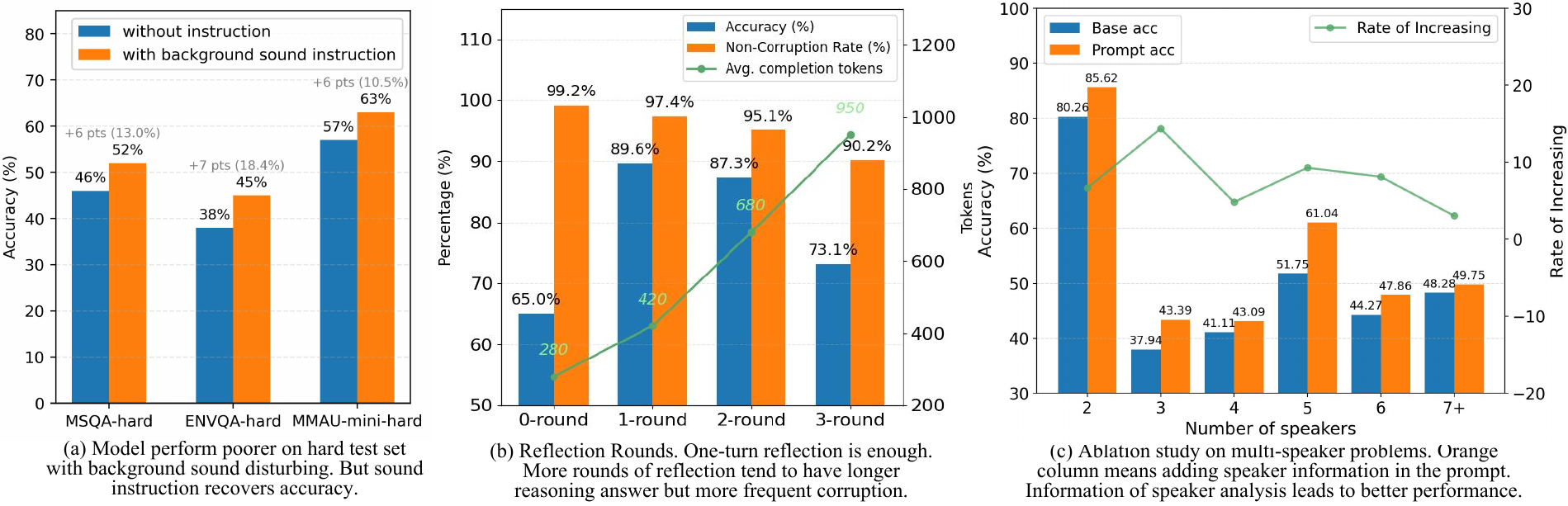}
\caption{Comparison between different audio situations. }
\label{fig:o1}
\end{figure*}

\subsection{Main Results}

\begin{table*}[t]
\centering
\small
\caption{Performance on MMAU Test-mini, MMAU-Test~\citep{MMAU}, MMAR~\citep{MMAR}, and MMSU~\citep{wang2025mmsu}. We report accuracy (\%) on MMAU on the Sound, Music, and Speech subsets and their averages, and on MMAR and MMSU, where higher is better.}
\setlength{\tabcolsep}{3.5pt}
\label{tab:main_merged}
\begin{tabular}{lcccc|cccc|cc}
\toprule
\multirow{2}{*}{\textbf{Method}} &
\multicolumn{4}{c|}{\textbf{MMAU Test-mini} $\uparrow$} &
\multicolumn{4}{c|}{\textbf{MMAU-Test} $\uparrow$} &
\multirow{2}{*}{\textbf{MMAR} $\uparrow$} &
\multirow{2}{*}{\textbf{MMSU} $\uparrow$} \\
\cmidrule(lr){2-5} \cmidrule(lr){6-9}
& Sound & Music & Speech & Avg.
& Sound & Music & Speech & Avg.
& Avg. & Avg. \\
\midrule
Gemini 2.5 Flash       & 67.97 & 62.28 & 62.76 & 64.30 & 65.43    & 65.30    & 63.30    & 64.68    & \textbf{66.80} & --    \\
GPT-4o            & 61.56 & 56.29 & 66.37 & 61.40 & 56.27    & 55.27    & 67.20    & 59.58    & \underline{63.50} & 56.38    \\
\midrule
Audio-Flamingo-3  & \textbf{79.58} & \underline{73.95} & 66.37 & \textbf{73.30} & \underline{75.83}    & \textbf{74.47}    & 66.97    & \textbf{72.42}    & 58.50 & --    \\
OmniVinci         & 73.65 & \textbf{78.68} & \underline{66.97} & \underline{73.10} & 73.07    & \underline{73.57}    & \underline{68.17}    & \underline{71.60}    & 58.30 & --    \\
Qwen2.5-Omni-7B   & \underline{78.10} & 65.90 & \textbf{70.60} & 71.50 & \textbf{76.77}    & 67.33    & \textbf{68.90}    & 71.00    & 56.70 & \textbf{60.57}    \\
Qwen2-Audio-7B-Instruct       & 61.26 & 53.59 & 48.05 & 54.30 & 55.27 & 48.56 & 42.13 & 48.65 & 30.00 & 48.31 \\
+SFT              & 62.76 & 44.61 & 55.86 & 54.41 & 61.17 & 55.67 & 55.37 & 57.40 & 40.90 & 51.03 \\
+GRPO             & 68.17 & 61.38 & 60.66 & 63.40 & 67.27 & 61.23 & 62.70 & 63.73 & 45.40 & 53.27 \\
+GRPO +ExpCoT     & 75.07 & 58.98 & 63.66 & 65.90 & --    & --    & --    & --    & 48.20 & --    \\
\textbf{Ours (HyPeR)} & 75.67 & 62.27 & 64.26 & 67.40 & 73.57 & 61.40 & 66.49 & 67.15 & 55.50 & \underline{56.38} \\
\midrule
Audio-CoT         & 62.16 & 55.99 & 56.16 & 58.10 & --    & --    & --    & --    & 31.67 & --    \\
Audio-Reasoner    & 60.06 & 64.30 & 60.70 & 61.71 & 61.56 & 55.99 & 53.45 & 57.00 & 36.71 & 35.51 \\
Audio-Thinker     & 76.88 & 62.87 & 64.26 & 68.00 & 75.13 & 61.83 & 67.03 & 67.90 & 52.00 & --    \\
\bottomrule
\end{tabular}
\end{table*}

We evaluate HyPeR against multiple LALMs on MMAU Test-mini and MMAR. As shown in Table~\ref{tab:main_merged}, our method achieves performance competitive with large-scale models on complex audio understanding tasks, particularly in speech.

\textbf{RL vs. SFT}  While GRPO without reasoning (No-Think) improves accuracy, the most substantial gains occur when combining Explicit Perceptual Traces (Stage I) with Implicit Latent Computation (Stage II). HyPeR offsets the domain shift observed in the Music subset during SFT, suggesting that RL helps the model adapt its perceptual boundaries to diverse acoustic scenes.

\textbf{PAUSE mechanism works.} The implicit reasoning enabled by \texttt{<PAUSE>} tokens during ambiguous acoustic phases is particularly effective in complex audio environments, especially on naturally occurring mixed-modality audio (MMAR +25.5). Notably, it improves the Music subset, offsetting the bad performance of just finetuning. More detailed analyses are provided in Appendix \ref{sec:PAUSE}.

\subsection{Ablation Study}

\subsubsection{Robustness to ENV and Multi-Speaker}
\paragraph{Background Sound} 
As shown in Fig.~\ref{fig:o1}(a), we evaluate that once the model is informed of background sound (one component of the prompt), it can correctly detect if that “noise” is unrelated to the dialogue content.
The introduction of background sound in the original audio leads to measurable degradation of zero-shot performance.
However, this drop is substantially mitigated while explicit ``ignore background sound" prompts are provided.
This validates that our reflection step improves accuracy.
In Fig.~\ref{fig:o1}(b), we further compare the effect of varying numbers of reflection turns, moving from 0 to 1 round, which yields a large accuracy enhancement. However, adding more rounds leads to “overthinking” and worse results, suggesting that longer reasoning is unnecessary.

\paragraph{Multi Speakers} Overall, recognizing the environment sound improves accuracy, which is consistently beneficial across all speaker counts. The base model is strong with 2 speakers (80.26\%), but drops sharply with more speakers. This pattern matches the intuition that more speakers introduce attribution and coreference errors. For 7+ speakers, the improvement is modest, suggesting that richer cues (e.g., explicit diarization tags, role summaries, or brief scene summaries) are likely needed.

\subsubsection{Reward Function}
As shown in Table~\ref{tab:ablation_reward}, we compare HyPeR and GRPO without Consistency Reward and length shaping respectively. The results demonstrate that the consistency reward ensures the model's logic is strictly grounded in the ASR and environment sound, leading to a 4.2\% gain in overall reliability.

\begin{table}[ht] \centering \small 
\caption{Ablation of rewards of Accuracy (Acc.) and Consistency (Cons.) on PAQA test dataset.} \label{tab:ablation_reward} 
\begin{tabular}{lcc} \toprule 
\textbf{Config} & \textbf{Acc.} & \textbf{Con.} \\ \midrule 
Full Reward (HyPeR) & \textbf{68.4} & \textbf{91.2} \\
w/o Consistency Reward ($\mathcal{R}_{\text{con}}$) & 64.2 & 78.5 \\ 
w/o Length Shaping ($\mathcal{R}_{\text{len}}$) & 67.1 & 89.4 \\ \bottomrule 
\end{tabular} 
\end{table}

\subsubsection{Do PAUSE Tokens Enable Latent Reasoning in Audio?}

To verify that PAUSE enables genuine latent reasoning rather than simply increasing decoding length, we analyze the evolution of the model's top-layer hidden states $h_{t}$ during the PAUSE phase by tracking two metrics across PAUSE indices $i$: (1) \textbf{Cosine Similarity to Answer} $\text{cos}(h_{PAUSE, i}, h_{ans})$, measuring how much the representation aligns with the final correct output; and (2) \textbf{Step-wise Displacement} $\|\Delta h\| = \|h_{i} - h_{i-1}\|$, quantifying the magnitude of state updates. As shown in Table~\ref{tab:PAUSE_stats}, the displacement $\|\Delta h\|$ remains significantly above zero, confirming that the hidden states are undergoing active transformation rather than staying stagnant. While initial PAUSEs may involve exploratory shifts, the trajectory eventually converges towards the answer embedding, suggesting that PAUSE serves as an adaptive latent computation mechanism rather than a redundant decoding delay. Representative success and failure cases are presented in Appendix \ref{sec:case}.

\begin{table}[ht]
\centering
\caption{Evolution of hidden states across sequential PAUSE tokens (averaged over 100 samples). Trigger Freq. indicates the proportion of samples in which the 1st, 2nd, or 3rd PAUSE is triggered.}
\label{tab:PAUSE_stats}
\resizebox{7.8cm}{!}{
\begin{tabular}{lcccc}
\toprule
Metric/PAUSE Token &  \#1 &  \#2 &  \#3 & Final Ans \\
\midrule
Avg. Cos-Sim to Ans & 0.47 & 0.51 & 0.62 & 0.73 \\
State Displacement $\|\Delta h\|$ & - & 336.2 & 324.8 & 338.5 \\
Trigger Freq. (per sample) & 1.00 & 0.78 & 0.45 & - \\
\bottomrule
\end{tabular}}
\end{table}

\section{Conclusion}

In this paper, we argue that improving audio understanding requires the base model to have audio grounding. Based on Auditory Scene Analysis, we focus on verifiable acoustic evidence and first introduce PAQA, a dataset that implements a layered decoupling strategy to separate speech from environmental interference and resolve multi-speaker attribution. Building upon this, we proposed HyPeR, a hybrid framework that unifies explicit perceptual reflections with implicit latent reasoning with GRPO-based <PAUSE> tokens. Experiments demonstrate that HyPeR significantly reduces perceptual errors and improves reasoning ability with evidence-constrained acoustic grounding.

\section*{Limitations}
Despite the significant improvements achieved by HyPeR and PAQA, several limitations remain to be addressed in future work: First, the introduction of the \texttt{<PAUSE>} token mechanism inevitably increases both training and inference latency.
Although our proposed Abort mechanism partially mitigates this, finding an balance between reasoning depth and real-time responsiveness remains a significant challenge.
Future work will explore more efficient latent reasoning techniques to minimize latency without sacrificing the robustness of audio grounding. In addition, while HyPeR performs well on several reasoning settings, it is weaker on some broader audio-language benchmarks, and stronger comparisons against more recent baselines would further clarify where the method is most effective.
Finally, PAQA is designed to emphasize perception-grounded reasoning, but it remains limited in scale and domain coverage.
Its construction involves strict structured curation and quality control.

\section*{Ethical Considerations}
Regarding Data Privacy, all audio samples in the PAQA dataset are derived from publicly available sources with permissive licenses, and any potentially sensitive speech content has been manually screened and anonymized to protect individual privacy. The license of MUSAN is CC\_BY 4.0, which permits free use for academic research and modification, and we have cited the work. 

\section*{Acknowledgments}
The research is supported by the AI for Science Program, Shanghai Municipal Commission of Economy and Informatization (Grant No. 2025-GZL-RGZN-BTBX-02028). The project’s computational resources are partially supported by CFFF platform of Fudan University.

\bibliography{custom}

@article{Guo2025DeepSeekR1,
  title={Deepseek-r1: Incentivizing reasoning capability in llms via reinforcement learning},
  author={Guo, Daya and Yang, Dejian and Zhang, Haowei and Song, Junxiao and Zhang, Ruoyu and Xu, Runxin and Zhu, Qihao and Ma, Shirong and Wang, Peiyi and Bi, Xiao and others},
    year={2025},
    eprint={2501.12948},
    archivePrefix={arXiv},
    primaryClass={cs.CL},
    url={https://arxiv.org/abs/2501.12948}, 
}

@misc{goyal2023think,
      title={Think before you speak: Training Language Models With Pause Tokens}, 
      author={Sachin Goyal and Ziwei Ji and Ankit Singh Rawat and Aditya Krishna Menon and Sanjiv Kumar and Vaishnavh Nagarajan},
      year={2024},
      eprint={2310.02226},
      archivePrefix={arXiv},
      primaryClass={cs.CL},
      url={https://arxiv.org/abs/2310.02226}, 
}

@article{shi2026qwen3,
  title={Qwen3-ASR Technical Report},
  author={Shi, Xian and Wang, Xiong and Guo, Zhifang and Wang, Yongqi and Zhang, Pei and Zhang, Xinyu and Guo, Zishan and Hao, Hongkun and Xi, Yu and Yang, Baosong and others},
  journal={arXiv preprint arXiv:2601.21337},
  year={2026}
}

@inproceedings{Poria2019,
    title = "{MELD}: A Multimodal Multi-Party Dataset for Emotion Recognition in Conversations",
    author = "Poria, Soujanya  and
      Hazarika, Devamanyu  and
      Majumder, Navonil  and
      Naik, Gautam  and
      Cambria, Erik  and
      Mihalcea, Rada",
    editor = "Korhonen, Anna  and
      Traum, David  and
      M{\`a}rquez, Llu{\'i}s",
    booktitle = "Proceedings of the 57th Annual Meeting of the Association for Computational Linguistics",
    month = jul,
    year = "2019",
    address = "Florence, Italy",
    publisher = "Association for Computational Linguistics",
    url = "https://aclanthology.org/P19-1050/",
    doi = "10.18653/v1/P19-1050",
    pages = "527--536"
}

@misc{wang2020covost,
      title={CoVoST 2 and Massively Multilingual Speech-to-Text Translation}, 
      author={Changhan Wang and Anne Wu and Juan Pino},
      year={2020},
      eprint={2007.10310},
      archivePrefix={arXiv},
      primaryClass={cs.CL},
      url={https://arxiv.org/abs/2007.10310}, 
}

@misc{Huang2025VisionR1,
      title={Vision-R1: Incentivizing Reasoning Capability in Multimodal Large Language Models}, 
      author={Wenxuan Huang and Bohan Jia and Zijie Zhai and Shaosheng Cao and Zheyu Ye and Fei Zhao and Zhe Xu and Yao Hu and Shaohui Lin},
      year={2025},
      eprint={2503.06749},
      archivePrefix={arXiv},
      primaryClass={cs.CV},
      url={https://arxiv.org/abs/2503.06749}, 
}

@misc{wang2025deepthink,
      title={Deep Think with Confidence}, 
      author={Yichao Fu and Xuewei Wang and Yuandong Tian and Jiawei Zhao},
      year={2025},
      eprint={2508.15260},
      archivePrefix={arXiv},
      primaryClass={cs.LG},
      url={https://arxiv.org/abs/2508.15260}, 
}

@misc{Feng2025VideoR1,
      title={Video-R1: Reinforcing Video Reasoning in MLLMs}, 
      author={Kaituo Feng and Kaixiong Gong and Bohao Li and Zonghao Guo and Yibing Wang and Tianshuo Peng and Junfei Wu and Xiaoying Zhang and Benyou Wang and Xiangyu Yue},
      year={2025},
      eprint={2503.21776},
      archivePrefix={arXiv},
      primaryClass={cs.CV},
      url={https://arxiv.org/abs/2503.21776}, 
}

@misc{AudioFlamingo,
      title={Audio Flamingo: A Novel Audio Language Model with Few-Shot Learning and Dialogue Abilities}, 
      author={Zhifeng Kong and Arushi Goel and Rohan Badlani and Wei Ping and Rafael Valle and Bryan Catanzaro},
      year={2024},
      eprint={2402.01831},
      archivePrefix={arXiv},
      primaryClass={cs.SD},
      url={https://arxiv.org/abs/2402.01831}, 
}

@misc{snyder2015musan,
      title={MUSAN: A Music, Speech, and Noise Corpus}, 
      author={David Snyder and Guoguo Chen and Daniel Povey},
      year={2015},
      eprint={1510.08484},
      archivePrefix={arXiv},
      primaryClass={cs.SD},
      url={https://arxiv.org/abs/1510.08484}, 
}

@misc{Chen2025AudioFlamingo3,
      title={Audio Flamingo 3: Advancing Audio Intelligence with Fully Open Large Audio Language Models}, 
      author={Arushi Goel and Sreyan Ghosh and Jaehyeon Kim and Sonal Kumar and Zhifeng Kong and Sang-gil Lee and Chao-Han Huck Yang and Ramani Duraiswami and Dinesh Manocha and Rafael Valle and Bryan Catanzaro},
      year={2025},
      eprint={2507.08128},
      archivePrefix={arXiv},
      primaryClass={cs.SD},
      url={https://arxiv.org/abs/2507.08128}, 
}

@misc{SALMONN,
      title={SALMONN: Towards Generic Hearing Abilities for Large Language Models}, 
      author={Changli Tang and Wenyi Yu and Guangzhi Sun and Xianzhao Chen and Tian Tan and Wei Li and Lu Lu and Zejun Ma and Chao Zhang},
      year={2024},
      eprint={2310.13289},
      archivePrefix={arXiv},
      primaryClass={cs.SD},
      url={https://arxiv.org/abs/2310.13289}, 
}

@misc{AudioCoT,
      title={Audio-CoT: Exploring Chain-of-Thought Reasoning in Large Audio Language Model}, 
      author={Ziyang Ma and Zhuo Chen and Yuping Wang and Eng Siong Chng and Xie Chen},
      year={2025},
      eprint={2501.07246},
      archivePrefix={arXiv},
      primaryClass={cs.SD},
      url={https://arxiv.org/abs/2501.07246}, 
}

@misc{AudioReasoner,
      title={Audio-Reasoner: Improving Reasoning Capability in Large Audio Language Models}, 
      author={Zhifei Xie and Mingbao Lin and Zihang Liu and Pengcheng Wu and Shuicheng Yan and Chunyan Miao},
      year={2025},
      eprint={2503.02318},
      archivePrefix={arXiv},
      primaryClass={cs.SD},
      url={https://arxiv.org/abs/2503.02318}, 
}

@misc{SARI,
      title={SARI: Structured Audio Reasoning via Curriculum-Guided Reinforcement Learning}, 
      author={Cheng Wen and Tingwei Guo and Shuaijiang Zhao and Wei Zou and Xiangang Li},
      year={2025},
      eprint={2504.15900},
      archivePrefix={arXiv},
      primaryClass={cs.CL},
      url={https://arxiv.org/abs/2504.15900}, 
}

@misc{Shao2024DeepSeekMathGRPO,
      title={DeepSeekMath: Pushing the Limits of Mathematical Reasoning in Open Language Models}, 
      author={Zhihong Shao and Peiyi Wang and Qihao Zhu and Runxin Xu and Junxiao Song and Xiao Bi and Haowei Zhang and Mingchuan Zhang and Y. K. Li and Y. Wu and Daya Guo},
      year={2024},
      eprint={2402.03300},
      archivePrefix={arXiv},
      primaryClass={cs.CL},
      url={https://arxiv.org/abs/2402.03300}, 
}

@misc{AudioThinker,
      title={Audio-Thinker: Guiding Audio Language Model When and How to Think via Reinforcement Learning}, 
      author={Shu Wu and Chenxing Li and Wenfu Wang and Hao Zhang and Hualei Wang and Meng Yu and Dong Yu},
      year={2025},
      eprint={2508.08039},
      archivePrefix={arXiv},
      primaryClass={cs.SD},
      url={https://arxiv.org/abs/2508.08039}, 
}

@article{R1AQA,
  title={Reinforcement Learning Outperforms Supervised Fine-Tuning: A Case Study on Audio Question Answering},
  author={Li, Gang and Liu, Jizhong and Dinkel, Heinrich and Niu, Yadong and Zhang, Junbo and Luan, Jian},
  journal={arXiv preprint arXiv:2503.11197},
  year={2025},
  url={https://github.com/xiaomi-research/r1-aqa; https://huggingface.co/mispeech/r1-aqa}
}

@inproceedings{shinn2023reflexion,
  title     = {Reflexion: Language Agents with Verbal Reinforcement Learning},
  author    = {Shinn, Noah and Cassano, Federico and Berman, Edward and Gopinath, Ashwin and Narasimhan, Karthik and Yao, Shunyu},
  booktitle = {Advances in Neural Information Processing Systems (NeurIPS)},
  year      = {2023},
  url       = {https://arxiv.org/abs/2303.11366}
}

@misc{madaan2023selfrefine,
      title={Self-Refine: Iterative Refinement with Self-Feedback}, 
      author={Aman Madaan and Niket Tandon and Prakhar Gupta and Skyler Hallinan and Luyu Gao and Sarah Wiegreffe and Uri Alon and Nouha Dziri and Shrimai Prabhumoye and Yiming Yang and Shashank Gupta and Bodhisattwa Prasad Majumder and Katherine Hermann and Sean Welleck and Amir Yazdanbakhsh and Peter Clark},
      year={2023},
      eprint={2303.17651},
      archivePrefix={arXiv},
      primaryClass={cs.CL},
      url={https://arxiv.org/abs/2303.17651}, 
}

@misc{wang2023selfconsistency,
      title={Self-Consistency Improves Chain of Thought Reasoning in Language Models}, 
      author={Xuezhi Wang and Jason Wei and Dale Schuurmans and Quoc Le and Ed Chi and Sharan Narang and Aakanksha Chowdhery and Denny Zhou},
      year={2023},
      eprint={2203.11171},
      archivePrefix={arXiv},
      primaryClass={cs.CL},
      url={https://arxiv.org/abs/2203.11171}, 
}

@misc{OmniR1,
      title={Omni-R1: Reinforcement Learning for Omnimodal Reasoning via Two-System Collaboration}, 
      author={Hao Zhong and Muzhi Zhu and Zongze Du and Zheng Huang and Canyu Zhao and Mingyu Liu and Wen Wang and Hao Chen and Chunhua Shen},
      year={2025},
      eprint={2505.20256},
      archivePrefix={arXiv},
      primaryClass={cs.CV},
      url={https://arxiv.org/abs/2505.20256}, 
}

@misc{MMAU,
      title={MMAU: A Massive Multi-Task Audio Understanding and Reasoning Benchmark}, 
      author={S Sakshi and Utkarsh Tyagi and Sonal Kumar and Ashish Seth and Ramaneswaran Selvakumar and Oriol Nieto and Ramani Duraiswami and Sreyan Ghosh and Dinesh Manocha},
      year={2024},
      eprint={2410.19168},
      archivePrefix={arXiv},
      primaryClass={eess.AS},
      url={https://arxiv.org/abs/2410.19168}, 
}

@misc{MMAR,
      title={MMAR: A Challenging Benchmark for Deep Reasoning in Speech, Audio, Music, and Their Mix}, 
      author={Ziyang Ma and Yinghao Ma and Yanqiao Zhu and Chen Yang and Yi-Wen Chao and Ruiyang Xu and Wenxi Chen and Yuanzhe Chen and Zhuo Chen and Jian Cong and Kai Li and Keliang Li and Siyou Li and Xinfeng Li and Xiquan Li and Zheng Lian and Yuzhe Liang and Minghao Liu and Zhikang Niu and Tianrui Wang and Yuping Wang and Yuxuan Wang and Yihao Wu and Guanrou Yang and Jianwei Yu and Ruibin Yuan and Zhisheng Zheng and Ziya Zhou and Haina Zhu and Wei Xue and Emmanouil Benetos and Kai Yu and Eng-Siong Chng and Xie Chen},
      year={2025},
      eprint={2505.13032},
      archivePrefix={arXiv},
      primaryClass={cs.SD},
      url={https://arxiv.org/abs/2505.13032}, 
}

@inproceedings{yang2022avqa,
  title={AVQA: A Dataset for Audio-Visual Question Answering on Videos},
  author={Yang, Pinci and Wang, Xin and Duan, Xuguang and Chen, Hong and Hou, Runze and Jin, Cong and Zhu, Wenwu},
  booktitle={Proceedings of the 30th ACM International Conference on Multimedia},
  pages={3480--3491},
  year={2022}
}

@misc{ACT,
      title={Adaptive Computation Time for Recurrent Neural Networks}, 
      author={Alex Graves},
      year={2017},
      eprint={1603.08983},
      archivePrefix={arXiv},
      primaryClass={cs.NE},
      url={https://arxiv.org/abs/1603.08983}, 
}

@inproceedings{PonderNet,
  author    = {Andrea Banino and Samuel Ritter and others},
  title     = {PonderNet: Learning to Ponder},
  booktitle = {ICML},
  year      = {2021}
}

@article{Qwen2Audio,
  title={Qwen2-Audio Technical Report},
  author={Chu, Yunfei and Xu, Jin and Yang, Qian and Wei, Haojie and Wei, Xipin and Guo, Zhifang and Leng, Yichong and Lv, Yuanjun and He, Jinzheng and Lin, Junyang and Zhou, Chang and Zhou, Jingren},
  year={2024},
      eprint={2407.10759},
      archivePrefix={arXiv},
      primaryClass={eess.AS},
      url={https://arxiv.org/abs/2407.10759}, 
}

@misc{SearchR1,
      title={Search-R1: Training LLMs to Reason and Leverage Search Engines with Reinforcement Learning}, 
      author={Bowen Jin and Hansi Zeng and Zhenrui Yue and Jinsung Yoon and Sercan Arik and Dong Wang and Hamed Zamani and Jiawei Han},
      year={2025},
      eprint={2503.09516},
      archivePrefix={arXiv},
      primaryClass={cs.CL},
      url={https://arxiv.org/abs/2503.09516}, 
}

@misc{GPT4oAudioPreview,
  author       = {OpenAI},
  title        = {GPT-4o Audio Model (gpt-4o-audio-preview) | OpenAI API Documentation},
  howpublished = {\url{https://platform.openai.com/docs/models/gpt-4o-audio-preview}},
  note         = {Accessed 2025-12-22}
}

@misc{Gemini25,
  author       = {Kavukcuoglu, Koray},
  title        = {Gemini 2.5: Our most intelligent AI model},
  year         = {2025},
  month        = mar,
  howpublished = {\url{https://blog.google/technology/google-deepmind/gemini-model-thinking-updates-march-2025/}},
  note         = {Accessed 2025-12-22}
}

@article{TwS,
  title={Thinking with Sound: Audio Chain-of-Thought Enables Multimodal Reasoning in Large Audio-Language Models},
  author={Xiong, Zhen and Cai, Yujun and Li, Zhecheng and Yuan, Junsong and Wang, Yiwei},
  journal={arXiv preprint arXiv:2509.21749},
  year={2025}
}

@article{ghosh2024gama,
  title={Gama: A large audio-language model with advanced audio understanding and complex reasoning abilities},
  author={Ghosh, Sreyan and Kumar, Sonal and Seth, Ashish and Evuru, Chandra Kiran Reddy and Tyagi, Utkarsh and Sakshi, S and Nieto, Oriol and Duraiswami, Ramani and Manocha, Dinesh},
  journal={arXiv preprint arXiv:2406.11768},
  year={2024}
}

@article{michelsanti2021overview,
  title={An overview of deep-learning-based audio-visual speech enhancement and separation},
  author={Michelsanti, Daniel and Tan, Zheng-Hua and Zhang, Shi-Xiong and Xu, Yong and Yu, Meng and Yu, Dong and Jensen, Jesper},
  journal={IEEE/ACM Transactions on Audio, Speech, and Language Processing},
  volume={29},
  pages={1368--1396},
  year={2021},
  publisher={IEEE}
}

@article{pass1yy,
  title={Does reinforcement learning really incentivize reasoning capacity in llms beyond the base model?},
  author={Yue, Yang and Chen, Zhiqi and Lu, Rui and Zhao, Andrew and Wang, Zhaokai and Song, Shiji and Huang, Gao},
  journal={arXiv preprint arXiv:2504.13837},
  year={2025}
}

@book{ASAbregman1994auditory,
  title={Auditory scene analysis: The perceptual organization of sound},
  author={Bregman, Albert S},
  year={1994},
  publisher={MIT press}
}

@article{fonseca2021fsd50k,
  title={Fsd50k: an open dataset of human-labeled sound events},
  author={Fonseca, Eduardo and Favory, Xavier and Pons, Jordi and Font, Frederic and Serra, Xavier},
  journal={IEEE/ACM Transactions on Audio, Speech, and Language Processing},
  volume={30},
  pages={829--852},
  year={2021},
  publisher={IEEE}
}

@inproceedings{xu2021text,
  title={Text-to-audio grounding: Building correspondence between captions and sound events},
  author={Xu, Xuenan and Dinkel, Heinrich and Wu, Mengyue and Yu, Kai},
  booktitle={ICASSP 2021-2021 IEEE International Conference on Acoustics, Speech and Signal Processing (ICASSP)},
  pages={606--610},
  year={2021},
  organization={IEEE}
}

@inproceedings{kuan2025can,
  title={Can large audio-language models truly hear? tackling hallucinations with multi-task assessment and stepwise audio reasoning},
  author={Kuan, Chun-Yi and Lee, Hung-yi},
  booktitle={ICASSP 2025-2025 IEEE International Conference on Acoustics, Speech and Signal Processing (ICASSP)},
  pages={1--5},
  year={2025},
  organization={IEEE}
}

@inproceedings{elizalde2023clap,
  title={Clap learning audio concepts from natural language supervision},
  author={Elizalde, Benjamin and Deshmukh, Soham and Al Ismail, Mahmoud and Wang, Huaming},
  booktitle={ICASSP 2023-2023 IEEE International Conference on Acoustics, Speech and Signal Processing (ICASSP)},
  pages={1--5},
  year={2023},
  organization={IEEE}
}

@article{niizumi2024m2d,
  title={M2D-CLAP: Masked modeling duo meets clap for learning general-purpose audio-language representation},
  author={Niizumi, Daisuke and Takeuchi, Daiki and Ohishi, Yasunori and Harada, Noboru and Yasuda, Masahiro and Tsubaki, Shunsuke and Imoto, Keisuke},
  journal={arXiv preprint arXiv:2406.02032},
  year={2024}
}

@inproceedings{elizalde2024natural,
  title={Natural language supervision for general-purpose audio representations},
  author={Elizalde, Benjamin and Deshmukh, Soham and Wang, Huaming},
  booktitle={ICASSP 2024-2024 IEEE International Conference on Acoustics, Speech and Signal Processing (ICASSP)},
  pages={336--340},
  year={2024},
  organization={IEEE}
}

@inproceedings{ghosh2025reclap,
  title={Reclap: Improving zero shot audio classification by describing sounds},
  author={Ghosh, Sreyan and Kumar, Sonal and Evuru, Chandra Kiran Reddy and Nieto, Oriol and Duraiswami, Ramani and Manocha, Dinesh},
  booktitle={ICASSP 2025-2025 IEEE International Conference on Acoustics, Speech and Signal Processing (ICASSP)},
  pages={1--5},
  year={2025},
  organization={IEEE}
}

@article{jaech2024openai,
  title={Openai o1 system card},
  author={Jaech, Aaron and Kalai, Adam and Lerer, Adam and Richardson, Adam and El-Kishky, Ahmed and Low, Aiden and Helyar, Alec and Madry, Aleksander and Beutel, Alex and Carney, Alex and others},
  journal={arXiv preprint arXiv:2412.16720},
  year={2024}
}

@article{comanici2025gemini,
  title={Gemini 2.5: Pushing the frontier with advanced reasoning, multimodality, long context, and next generation agentic capabilities},
  author={Comanici, Gheorghe and Bieber, Eric and Schaekermann, Mike and Pasupat, Ice and Sachdeva, Noveen and Dhillon, Inderjit and Blistein, Marcel and Ram, Ori and Zhang, Dan and Rosen, Evan and others},
  journal={arXiv preprint arXiv:2507.06261},
  year={2025}
}

@article{xu2025qwen2,
  title={Qwen2. 5-omni technical report},
  author={Xu, Jin and Guo, Zhifang and He, Jinzheng and Hu, Hangrui and He, Ting and Bai, Shuai and Chen, Keqin and Wang, Jialin and Fan, Yang and Dang, Kai and others},
  journal={arXiv preprint arXiv:2503.20215},
  year={2025}
}

@article{arora2025training,
  title={Training language models to reason efficiently},
  author={Arora, Daman and Zanette, Andrea},
  journal={arXiv preprint arXiv:2502.04463},
  year={2025}
}

@article{kong2023universal,
  title={Universal source separation with weakly labelled data},
  author={Kong, Qiuqiang and Chen, Ke and Liu, Haohe and Du, Xingjian and Berg-Kirkpatrick, Taylor and Dubnov, Shlomo and Plumbley, Mark D},
  journal={arXiv preprint arXiv:2305.07447},
  year={2023}
}

@misc{xue2025hhcodec,
      title={HH-Codec: High Compression High-fidelity Discrete Neural Codec for Spoken Language Modeling}, 
      author={Rongkun Xue and Yazhe Niu and Shuai Hu and Zixin Yin and Yongqiang Yao and Jing Yang},
      year={2025},
      eprint={2507.18897},
      archivePrefix={arXiv},
      primaryClass={cs.SD},
      url={https://arxiv.org/abs/2507.18897}, 
}

@misc{lightrft,
  title={LightRFT},
  author={Niu, Yazhe and Pu, Yuan and Shi, Dongxing and Lu, Yudong and Xiong, Yingtong and Ge, Ruijun and Sun, Jiaxuan and Wan, Zunian and Zhang, Shaoang and others},
  publisher={GitHub},
  howpublished={\url{https://github.com/opendilab/LightRFT}},
  year={2025},
}

@article{wang2025mmsu,
  title={{MMSU}: A massive multi-task spoken language understanding and reasoning benchmark},
  author={Wang, Dingdong and Wu, Jincenzi and Li, Junan and Yang, Dongchao and Chen, Xueyuan and Zhang, Tianhua and Meng, Helen},
  journal={arXiv preprint arXiv:2506.04779},
  year={2025}
}

@article{omnivinci2025,
      title={OmniVinci: Enhancing Architecture and Data for Omni-Modal Understanding LLM},
      author={Hanrong Ye and Chao-Han Huck Yang and Arushi Goel and Wei Huang and Ligeng Zhu and Yuanhang Su and Sean Lin and An-Chieh Cheng and Zhen Wan and Jinchuan Tian and Yuming Lou and Dong Yang and Zhijian Liu and Yukang Chen and Ambrish Dantrey and Ehsan Jahangiri and Sreyan Ghosh and Daguang Xu and Ehsan Hosseini-Asl and Danial Mohseni Taheri and Vidya Murali and Sifei Liu and Jason Lu and Oluwatobi Olabiyi and Frank Wang and Rafael Valle and Bryan Catanzaro and Andrew Tao and Song Han and Jan Kautz and Hongxu Yin and Pavlo Molchanov},
      journal={arXiv},
      year={2025},
}

\appendix
\clearpage
\newpage
\label{sec:appendix}
\section{Dataset Construction and Quality Control}
\label{sec:dataAna}
\subsection{Synthetic Audio with Background Sound}
Following this, we further analyze erroneous predictions of Qwen2-Audio on the MMAU benchmark.

As shown in Fig.~\ref{fig:appASRanalysis}(b), we compare fine-tuning trajectories on the MSQA dataset with and without ASR-augmented data.
The results reveal that models trained with ASR supervision exhibit substantially longer response lengths, which we interpret as a proxy for deeper and more structured reasoning ability.
This finding suggests that integrating ASR data into training not only improves transcription accuracy but also enhances the reasoning capacity of models.
Therefore, in the first stage of fine-tuning, we deliberately incorporate the ASR-enriched data described in the previous section to further consolidate the model’s capability as a foundation for downstream reasoning.

Moreover, we processed the audio with MUSAN\citep{snyder2015musan}, which satisfies target 10 dB SNR, according to 
$$\mathrm{SNR}{\mathrm{dB}}=10\log{10}\left(\frac{P_s}{P_{n,\mathrm{scaled}}}\right)=10.$$
Let $P_s=\frac{1}{T}\sum_t s_t^2$ and $ P_n=\frac{1}{T}\sum_t n_t^2$. The background gain is
$$
k=\sqrt{\frac{P_s}{P_n \cdot 10^{\mathrm{SNR}_{\mathrm{dB}}/10}}}=\sqrt{\frac{P_s}{P_n \cdot 10}}.$$

\subsection{Audio Question-Answering with Multi Speakers}
We use the subset of Multi-Speaker Dataset in CoTA~\citep{AudioReasoner}, which is generated by TTS to navigate intricate speaker interactions. First, we generated diverse conversational texts with LLMs. Next, using timbres from LibriSpeech as prompts, we synthesized high-quality speech via the CosyVoice2 framework. Finally, these distinct speech samples were combined into a rich dataset.

\section{Prompt Templates}
\label{sec:prompt}

Starting from heterogeneous audio sources, including clean speech corpora, multi-speaker question answering data, and background/environmental sound collections, the framework first constructs the PAQA training set through hierarchical audio augmentation. This augmentation explicitly decomposes complex auditory scenes along two perceptual axes inspired by Auditory Scene Analysis (ASA): (1) Environment vs. Speech, which teaches the model to separate foreground linguistic content from background noise or music; and (2) Speaker vs. Speaker, to distinguish different speakers, track turn-taking, and recover speaker-specific evidence.

\begin{tcolorbox}[
  breakable,
  enhanced,
  title=A Sample Case of Our PAQA Data,%
  colback=gray!10,
  colframe=gray!70!black,
  arc=1mm,
  boxrule=1pt,
  pad at break=1mm,
  left=1.5mm,
  right=1.5mm,
  top=1.5mm,
  bottom=1mm
]
\small
\texttt{<THINK>}\newline
    \texttt{<PLANNING>}
        The user wants to understand the dynamic changes...
    \texttt{</PLANNING>}\newline
    \texttt{<CAPTION>}
        \texttt{<BGM>}Street noise; intermittent car horns.\texttt{</BGM>}
        \texttt{<SPEAKER>}A: adult F; B: adult M; alternating turns.\texttt{</SPEAKER>}
        \texttt{<ASR>}...\texttt{</ASR>}
        \texttt{<DESCRIPTION>}The audio clip predominantly features static noise.... similar to that of a detuned television or a device failing to receive a signal.\texttt{</DESCRIPTION>}
    \texttt{</CAPTION>}\newline
    \texttt{<REASONING>}
        1. Identify changes in Intensity (Volume):... struggling 
        to maintain a consistent output, adding to the impression of
        something malfunctioning or broken. 
    \texttt{</REASONING>}\newline
    \texttt{<SUMMARY>}
        The static noise in the audio is highly dynamic. ... leading 
        to a sense of disorder and instability.
    \texttt{</SUMMARY>}\newline
\texttt{</THINK>}\newline
\texttt{<RESPONSE>}
    The audio presents a static noise,... is one of energetic 
    chaos, preventing any possibility of calm or predictability.
\texttt{</RESPONSE>}\newline
\texttt{<REFLECT1>} Does "A" mention the cake, not B? Check turn 3.\texttt{</REFLECT1>}\newline
\texttt{<NEW\_RESPONSE>}A\texttt{</NEW\_RESPONSE>}\newline
\texttt{<REFLECT2>} Does "A" mention the cake, not B? Check turn 3.\texttt{</REFLECT2>}\newline
\texttt{<NEW\_RESPONSE>}B\texttt{</NEW\_RESPONSE>}
\end{tcolorbox}

Based on this decomposition, each training example is converted from a simple audio-question-answer pair into a grounded reasoning target containing verifiable acoustic evidence, such as background sound tags, raw speech transcripts, speaker attribution, intermediate reasoning traces, reflection, and the corrected final answer. In this way, the model is trained not only to answer the question, but also to expose why the answer is supported.

\begin{figure}[hb]
\centering
\includegraphics[width=\linewidth]{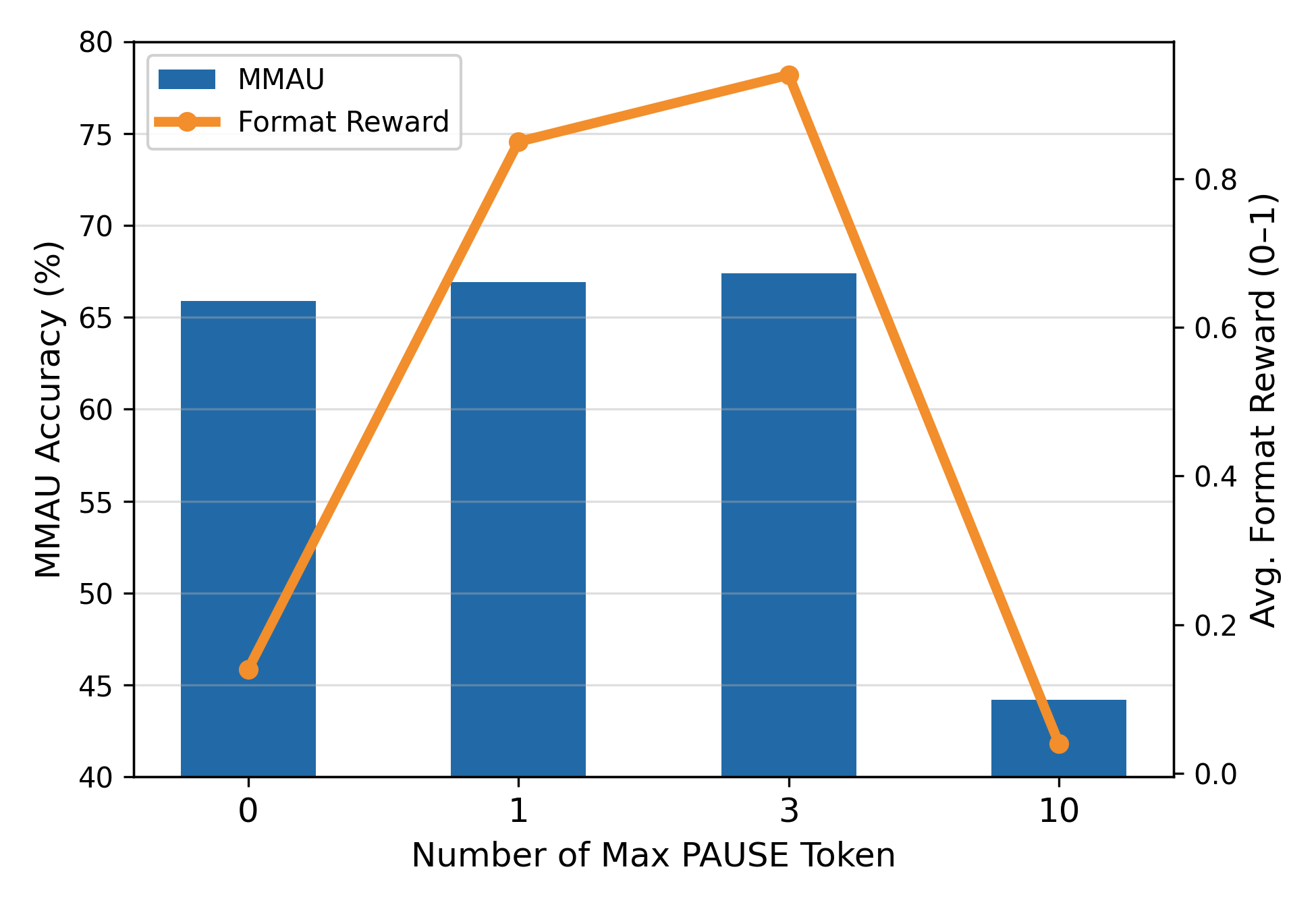}
\captionof{figure}{Abaltion study of \#\texttt{<PAUSE>} tokens. Set max PAUSE token as 1-3 is suitable.}
\label{fig:paqa_plot}
\end{figure}

\begin{tcolorbox}[
  breakable,
  enhanced,
  title=Prompt template of Refelection Sample,%
  colback=gray!10,
  colframe=gray!70!black,
  arc=1mm,
  boxrule=1pt,
  pad at break=1mm,
  left=1.5mm,
  right=1.5mm,
  top=1.5mm,
  bottom=2mm
]
\small%

After producing the \texttt{<RESPONSE>}, you must perform a structured self-reflection step.

1. Compare the \texttt{<RESPONSE>} with the overall task requirements and check for issues such as:
   - Missing or incomplete coverage of the audio content (did it stop too early? were some speakers/segments missed?).
   - Repetition or redundant phrasing that should be removed or marked clearly.
   - Speaker attribution or diarization errors (wrong speaker assignment, merged speakers, or split speakers).
   - Prosody/tone/intonation mistakes or overemphasis on irrelevant details.
   - Inconsistent reasoning or labels (final choice must align with the reasoning and context).
   - Overly simplistic or single-hypothesis reasoning when alternatives exist.

2. Inside \texttt{<REFLECT>}...\texttt{</REFLECT>}, explicitly list:
   - The problems found in \texttt{<RESPONSE>}.
   - The corrections or adjustments needed (without referencing or leaking the gold standard answer text).
   - Any uncertainties or low-confidence areas.

3. Then rewrite the improved answer inside \texttt{<FINAL\_ANSWER>}...\texttt{</FINAL\_ANSWER>}, ensuring:
   - All necessary content is covered.
   - No hallucinated details are added beyond the given \texttt{<CAPTION>}, \texttt{<ASR>}, and \texttt{<DESCRIPTION>}.
   - Speaker attributions and reasoning are consistent.
   - The final answer matches the reasoning and is labeled correctly with confidence if required.

Format strictly as:
\texttt{<REFLECT>}
[Your structured reflection here]
\texttt{</REFLECT>}

\texttt{<FINAL\_ANSWER>}
[Your corrected, high-quality final answer here]
\texttt{</FINAL\_ANSWER>}

Here is the original bad answer: {Turn0}
Here is the golden answer: {Golden\_Ans}

\end{tcolorbox}

\section{Data Statistics}
\label{app:dataAnalysis}
An illustrative example from the dataset is shown in Figure~\ref{fig:motivation}.
The dataset supports a broad range of tasks, including multi-speaker question answering, speech-to-text translation under noisy conditions, and environment-centric question answering. A comprehensive analysis of the final PAQA dataset is provided in Appendix~\ref{sec:dataAna}, while a statistical summary is presented in Table~\ref{tab:dataStat}.

\section{Additional Experimental Results}
\label{app:moreExp}
\subsection{Number of the PAUSE Tokens}
Excessive pausing negatively affects performance(see Figure~\ref{fig:paqa_plot}), suggesting that it is suitable to set max PAUSE token between 1 and 3.

\subsection{Results on the test set of PAQA}
We also evaluate on the test set of PAQA, on the category of multi-speaker and MELD~\citep{AudioReasoner}, HyPeR performs the best. The results is listed in Table. \ref{tab:main2}.

\begin{table}[htp]
\centering
\small
\setlength{\tabcolsep}{1pt}
\captionof{table}{Evaluation on the test set of PAQA. We use Qwen2-Audio-7B-Instruct as the base model. Our model performs best in each category.}
\begin{tabular}{lcccc}
\toprule
\multirow{2}{*}{Model} & \multicolumn{2}{c}{Multi-Speaker (hard)} & \multicolumn{2}{c}{BGM-rich Acc.}\\
 & Acc. $\uparrow$ & Con. $\uparrow$ & SNR=10 $\uparrow$ & SNR=5 $\uparrow$  \\
\midrule
Base      & 42.2 & 38.5 & 41.0 & 20.1 \\
+SFT              & 46.2 & 41.5 & 44.0 & 31.2 \\
+GRPO-NoThink     & 52.7 & 48.3 & 50.2 & 38.4 \\
+GRPO-ExpCoT      & 61.5 & 58.7 & 60.8 & 47.6 \\
\textbf{Ours} & \textbf{70.4} & \textbf{68.1} & \textbf{69.5} & \textbf{57.8} \\
\midrule
Audio-CoT        & 50.6 & 46.9 & 48.3 & 35.0 \\
Audio-Reasoner   & 56.8 & 52.7 & 55.9 & 41.8 \\
\bottomrule
\vspace{1pt}
\end{tabular}
\label{tab:main2}
\end{table}

Furthermore, under the challenging setting with background sound at SNR=5dB, a condition that considerably degrades most models, our HyPeR deteriorates the least, retaining state-of-the-art accuracy and consistency.
This resilience is attributed to its PAUSE-driven implicit reasoning and rewards aware of background sound/music.

\subsection{Proper Response Length after Latent Reasoning}
\label{sec:PAUSE}

Although PAUSE-based latent tokens improve training stability, they substantially increase training cost: increasing max\_PAUSE\_token from 1 to 3 roughly doubles training time. We therefore incorporate a length reward into the overall reward design and further analyze its effects in Sec.\ref{sec:lengthReward}. Overall, RL training is effective, but a noticeable performance drop often occurs around 200 steps. We attribute this instability to the length reward: during exploration, the model is encouraged to generate longer responses, but once the output exceeds ~600 tokens, a linear decay penalty applies. The policy then shifts toward shorter, often incomplete outputs, causing the format reward to collapse to 0 and the accuracy reward to drop to 0.5. But the training subsequently stabilizes, suggesting that the policy can adapt to this complex reward.

\subsection{Inference Efficiency}
For inference efficiency, our hybrid reasoning does not invoke multiple models or external modules at inference time. Instead, HyPeR is implemented by introducing a special PAUSE token into the decoder stream on the fine-tuned Qwen2-Audio-7B-Instruct backbone. During decoding, HyPeR may generate PAUSE tokens to perform latent reasoning steps, where the corresponding intermediate outputs are ignored (not fed back autoregressively), consistent with the “Ignore Output” mechanism described in our method. Therefore, the additional inference cost mainly comes from extra Transformer steps induced by PAUSE tokens.

To quantify this overhead, we evaluate models on one H200, with the batch size of 32, max\_new\_tokens of 2048. Results are listed in the Table \ref{tab:inferEff} below. Since the framework is inspired by ASA, we added a Cascaded ASA + LALM baseline, which first run ASA with a sound separation frontend called Universal Sound Separation~\cite{kong2023universal}, then ask the LALM (also finetuned for reasoning) to analyze each source and combine the responses.

\subsection{Ablation isolating Perception-Attention vs. Self-Correction}
To evaluate the effect of self-correction without enhanced perception, we construct a reflection-specific subset \(T'\) from the PAQA training set, containing 9,584 samples. On this subset, we compare \textbf{Qwen2-Audio-7B-Instruct} with \textbf{0} versus \textbf{1} reflection turn, without introducing any additional perception-side mechanism. We find that adding a single reflection turn improves accuracy from \textbf{49.17\%} (4,712/9,584) to \textbf{54.18\%} (5,193/9,584), yielding only a modest gain of \textbf{5.01}\%.

However, this improvement is accompanied by substantial prediction instability: \textbf{2,133} samples that are originally correct become incorrect after reflection. In contrast, when perception-aware mechanisms are enabled, the 1-turn accuracy increases to \textbf{89.50\%} (8,578/9,584), while the number of correct-to-incorrect cases drops sharply to only \textbf{11}. These results suggest that self-correction alone is not reliably effective, and its benefit critically depends on strong perception capabilities that stabilize the reflection process.

To isolate the effect of \textbf{Perception-Attention}, we evaluate the trained \textbf{HyPeR} model by retaining only the first answer generated in the \texttt{<Response>} stage, thereby removing the effect of self-correction at inference time. This \textbf{Perception Only} setting measures the contribution of improved perception in isolation. As shown in Table~\ref{tab:perception_only}, Perception Only already yields \textbf{63.20} average accuracy on MMAU-test-mini and \textbf{46.30} on MMAR, indicating that the proposed perception enhancement substantially improves the model's initial responses. The full HyPeR model further raises performance to \textbf{67.40} and \textbf{55.50}, respectively, showing that reflection contributes additional gains once a strong perception basis is established. Together, these results suggest that HyPeR benefits from both components, with Perception-Attention serving as the primary source of stable improvement.

\begin{table}[t]
\centering
\small
\caption{Ablation on the Perception-Attention component. ``Perception Only'' denotes using only the first answer from the \texttt{<Response>} stage, without self-correction at inference time.}
\label{tab:perception_only}
\setlength{\tabcolsep}{2pt}
\begin{tabular}{lccccc}
\toprule
\multirow{2}{*}{\textbf{Model}} & \multicolumn{4}{c}{\textbf{MMAU-Test mini}} & \multirow{2}{*}{\textbf{MMAR}} \\
\cmidrule(lr){2-5}
& \textbf{Sound} & \textbf{Music} & \textbf{Speech} & \textbf{Avg.} & \\
\midrule
Perception Only & 70.27 & 59.88 & 59.46 & 63.20 & 46.30 \\
\textbf{Ours (HyPeR)} & \textbf{75.67} & \textbf{62.27} & \textbf{64.26} & \textbf{67.40} & \textbf{55.50} \\
\bottomrule
\end{tabular}
\end{table}

\begin{table}[t]
\centering
\small
\caption{Comparison of performance and generation statistics on MMAU-test, where ``Len.'' denotes the completion length (\# of generated tokens), and ``Speed'' denotes the number of samples processed per second. HyPeR is configured with up to three PAUSE tokens, while all other methods are evaluated without PAUSE.}
\label{tab:mmautest_efficiency}
\footnotesize
\resizebox{7.9cm}{!}{
\begin{tabular}{lccc}
\toprule
\textbf{Method} & \textbf{MMAU} & \textbf{Len.} & \textbf{Speed} $\uparrow$ \\
\midrule
Audio-Reasoner              & 57.00         & 812 $\pm$ 356   & 5.82          \\
Qwen2-Audio-7B-Instruct     & 48.65         & 67 $\pm$ 5      & \textbf{8.16} \\
+SFT                        & 57.40         & 946 $\pm$ 402   & 3.77          \\
+GRPO                       & 63.73         & 612 $\pm$ 410   & 5.00          \\
USS+                        & 61.06         & 2200 $\pm$ 1100 & 1.51          \\
\textbf{Ours (HyPeR)}       & \textbf{67.15} & 781 $\pm$ 448   & 4.62          \\
\bottomrule
\end{tabular}
}
\label{tab:inferEff}
\end{table}

\begin{figure*}[htp]
\centering
\includegraphics[width=0.9\linewidth]{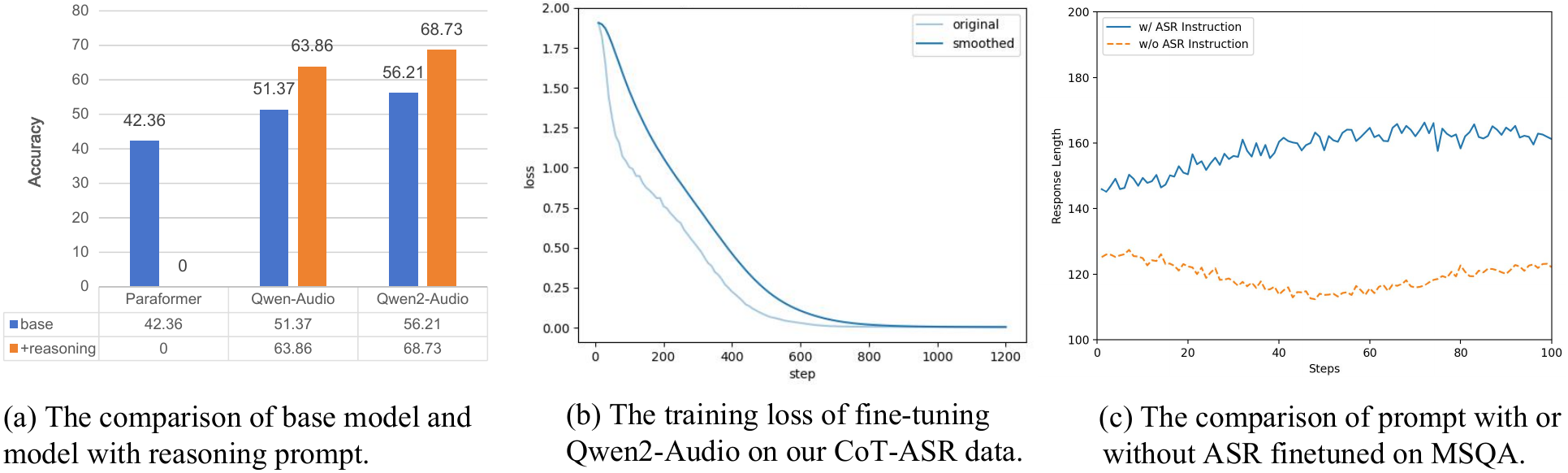}
\caption{Experiments on the Exploration of Good Audio Reasoning prompt.}
\label{fig:appASRanalysis}
\end{figure*}

\begin{table*}[!ht]
\centering
\caption{Dataset Source and Statistics. ``MS" means whether there are multi speakers in the audio.}
\label{tab:dataStat}
\resizebox{\linewidth}{!}{ 
\begin{tabular}{clccccc}
\toprule
\textbf{Dataset Source} & \textbf{Main Skills Learning} & \textbf{BGM Used} & \textbf{Quantity} & \textbf{Reflection} & \textbf{duration} & \textbf{MS}  \\ \midrule
Multi-Speaker \citep{AudioReasoner} & Multi-speaker Speech QA & Free Sound & 2.9k & 1.4k & 264 & $\checkmark$ \\ 
MELD \citep{Poria2019} & Speech Emotion QA & Sound Bible & 2.9k & 1.4k & 359 & $\checkmark$ \\ 
CoVoST2 \citep{wang2020covost} & Speech-to-Text Translation & No & 1.4k & No  & 72 & $\times$\\ 
\bottomrule
\end{tabular}}
\end{table*}

\section{Case Study}
\label{sec:case}
\begin{figure*}[t]
\centering
\includegraphics[width=1\linewidth]{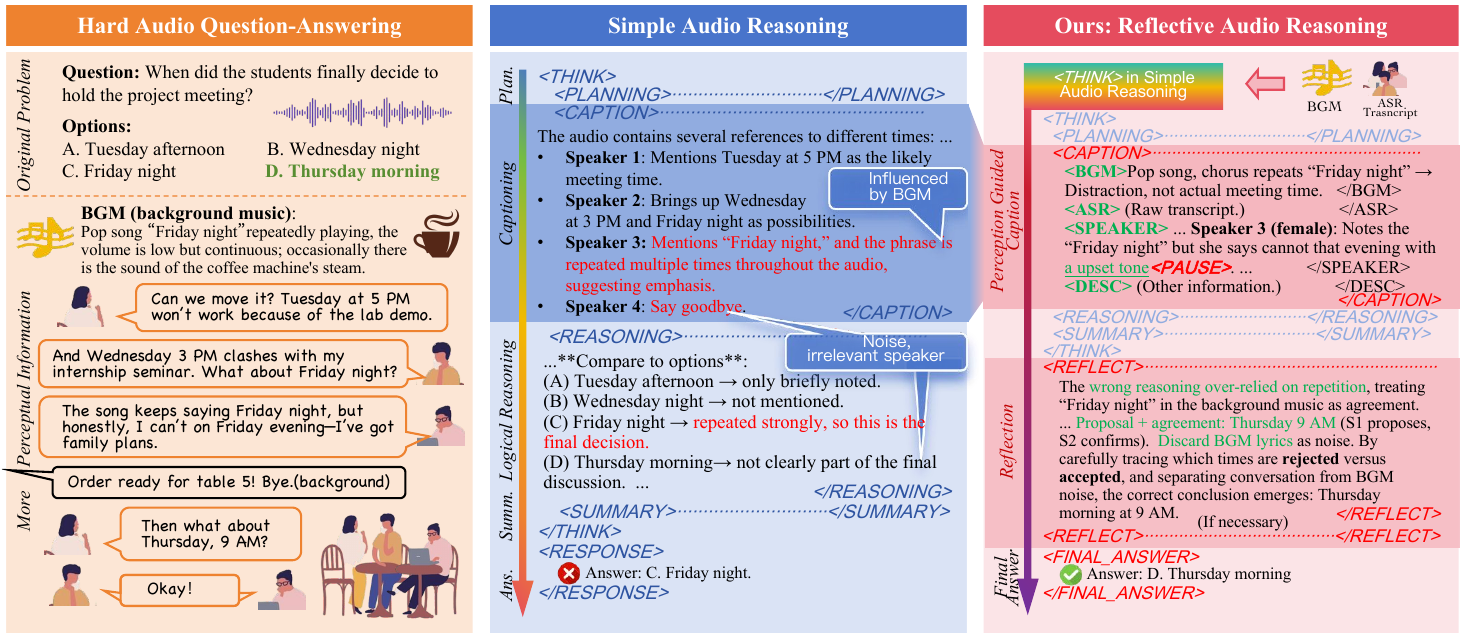}
\caption{Case study.}
\label{fig:motivation}
\end{figure*}

As shown in Fig.\ref{fig:motivation}, the case highlights two failure modes: perceptual misbinding and salience-driven rationale drift. The naive system exhibits this by prioritizing the surface frequency of "Friday" while overlooking its negated polarity and the logical flow of the proposal-to-confirmation sequence. Conversely, the reflective controller rectifies this by enforcing evidence typing (differentiating background sounds from linguistic turns) and ensuring dialogue-act alignment, ultimately restoring causal fidelity to the acoustic evidence.

\subsection{Explicit Reflection}
Original CoT concludes bed from generic cues, “piece A vs. B,” “bevel(ed) edge,” “don’t misassemble”—none of which uniquely identify a bed. This is a classic reasoning overreach: drawing a strong category conclusion from evidence that is common to many flat-pack builds. No category-specific anchor terms (e.g., headboard, shelf, tabletop, backrest) appear in the dialogue.

\begin{tcolorbox}[
enhanced,
title=Bad Case A,%
colback=gray!10,      
colframe=gray!70!black,
arc=1mm,%
boxrule=1pt,%
pad at break=0mm,
left=1mm,%
right=1mm,%
top=1mm,%
bottom=1mm]%
\small%
\texttt{<THINK>}\newline
\texttt{<PLANNING>}The user's question is: Based on the conversation, what are the speakers trying to assemble? I need to analyze the dialogue between the speakers to identify the item they are constructing. The audio provides the audio clip, and I should analyze it for details related to the assembly activity. \texttt{</PLANNING>}
\texttt{<CAPTION>} The audio features a conversation among three individuals in what sounds like a workspace. They are discussing an instruction manual and the assembly process.
- Speaker 1: Expresses frustration with the instruction manual, stating that it's difficult to distinguish pieces A and B.
- Speaker 2: Proposes that piece B has a slightly beveled edge, but concedes that it's similar.
- Speaker 3: Agrees that it looks like Bevel Edge and expresses hope that it's correct.
- Speaker 1: States it is Bevel Edge B, and declares they can complete the assembly.
- Speaker 2: Hope is right and doesn't want to misassemble again.
- Speaker 1: Confirms it is Bevel Edge B and asks everyone to take deep breaths and move on.\newline
The core task is assembling an item where they are currently putting together parts, such as bevel edges.
\texttt{</CAPTION>}\newline
\texttt{<REASONING>}\newline
1. Analyzing: The dialogue revolves around distinguishing parts (A and B) and confirming a beveled edge.\newline
2. Identifying the Assembly Activity: The key element is Bevel Edge, suggesting the speakers are assembling something with beveled sides.\newline
3. Evaluating Options:
  - (a) A table: While a table might require beveling, the central focus isn't just about assembling a table.
  - (b) A chair: Similarly, a chair might require beveling, but it isn't their primary focus.
  - (c) A bed: A bed typically features beveled edges, aligning perfectly with the conversation. This is the most likely scenario.
  - (d) A bookshelf: Bookshelf is less likely to have beveled edges.\newline
4. Concluding: Based on the bevel edges and the distinct focus on distinguishing them, it can be inferred that the speakers are assembling a bed.
\texttt{</REASONING>}\newline
\texttt{<SUMMARY>} The speakers are differentiated by focusing on the Bevel Edge assembly of a bed. By eliminating other less relevant options through reasoning, the final conclusion is that they are assembling a bed.
\texttt{</SUMMARY>}\newline
\texttt{</THINK>}
\end{tcolorbox}

Here is the reflection of this case.

\begin{tcolorbox}%
[
title=Reflection of Bad Case A,%
colback=gray!10,      
colframe=gray!70!black,
arc=1mm,%
boxrule=1pt,%
left=1mm,%
right=1mm,%
top=1mm,%
bottom=1mm]%
\small%
\texttt{<REFLECT>}
Evidence available from the dialogue is generic to many flat-pack builds: “piece A vs. B,” “bevel(ed) edge,” “don’t misassemble,” and “let’s move on.” No category-specific anchors are heard (e.g., for a bed: headboard/footboard/rails/slats; for a bookshelf: shelf/upright/back panel/bracket; for a table: tabletop/apron; for a chair: backrest/seat/armrest).  
Given absent anchors, the safest policy is to avoid over-commitment. If forced to choose among the four, “bookshelf” is the most compatible with beveled front edges on shelves and frequent A/B part distinctions during shelf panel assembly. Confidence remains low; with an “uncertain” option, I would state insufficient evidence.
\texttt{</REFLECT>}\newline
\texttt{<ANSWER>}Bookshelf.\texttt{</ANSWER>}

\end{tcolorbox}

\subsection{Multi Speakers Tracking}

This case tests how well the model can follow a conversation in a realistic, noisy meeting. The audio includes a constant low hum from an air conditioner (HVAC) in the background, and the speakers often interrupt each other. These challenges make it difficult to follow the flow of the conversation. The simple model's failure wasn't about hearing the words; it was about failing to understand the process of the discussion. The reflective model succeeded because it built a step-by-step understanding of the conversation's logic. It correctly handled the disagreements and changes, proving it can reason reliably even in a noisy and dynamic group setting.

\begin{tcolorbox}%
[
title=Bad Case B,%
colback=gray!10,      
colframe=gray!70!black,
arc=1mm,%
boxrule=1pt,%
left=1mm,%
right=1mm,%
top=1mm,%
bottom=1mm]%
\small%
[ASR excerpt] 
[S1] “Ship date is the 12th if QA passes.”
[S3] “QA won’t finish by the 12th.”
[S4] “Set the launch to the 15th.”
[S2] “Not the 5th—I said the 15th.”
[S1] “Agreed.”\newline
[Question] What is the final launch date?\newline
[Choices] A 5th · B 12th · C 13th · D 15th\newline
[Audio context] Meeting room; steady HVAC hum as BGM; frequent interruptions.\newline
[Ground truth] D\newline
[Simple model response] B \newline
[Reflection] \texttt{<REFLECT>}
S3 explicitly negates the initial “12th” suggestion, S4 introduces a new proposal “15th,” S2 clarifies that it is “15th” rather than “5th,” and S1 provides team confirmation; taken together these turns establish the final launch date as the 15th.
\end{tcolorbox}

\subsection{PAUSE Mechanism}

\begin{figure*}[t]
\centering
\includegraphics[width=1\linewidth]{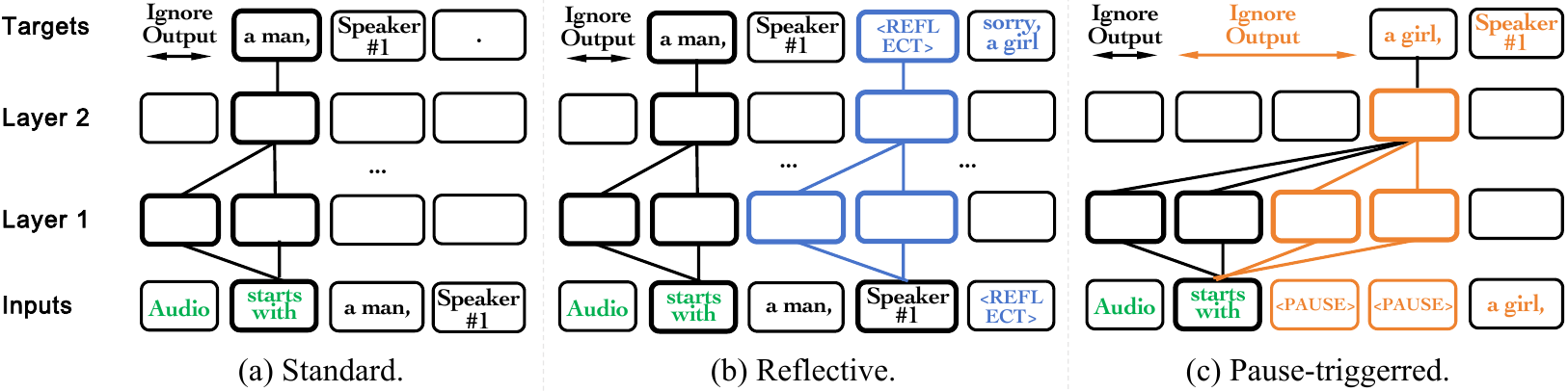}
\caption{Standard vs. reflective-inference vs. PAUSE-inference (and finetuning). Similar to \citet{goyal2023think}, the rounded squares denote one Transformer operation (a self-attention and MLP) in a 2-layer Transformer. Any Ignore Output denotes that during inference, the output token is not extracted and not fed back autoregressively.}
\label{fig:framwork}
\end{figure*}

This case tests the model's ability to understand meaning that isn't in the words themselves, but in the way they are spoken (the "tone," or prosody). The audio is a phone call with background noise from rain and static (line hiss). The challenge is to see if the model can spot sarcasm or reluctance, even when the words seem positive. Therefore, this case highlights a major flaw in models that only read text. They can be easily fooled when a speaker's tone changes the meaning of their words (like in sarcasm or politeness). The reflective model's success shows it can listen, not just read. It integrates how something is said with what is said to find the true meaning, which is crucial for understanding human conversation.

\begin{tcolorbox}%
[
title=Bad Case C,%
colback=gray!10,      
colframe=gray!70!black,
arc=1mm,%
boxrule=1pt,%
left=1mm,%
right=1mm,%
top=1mm,%
bottom=1mm]%
\small%
[ASR excerpt] 
[S1] “Can you present Friday?”
[S2] “Yeah, Friday… great.” (F0 downward; reduced energy; elongated final syllable)
[S2] “Actually, I’m not available Friday evening.”
[S1] “Then Thursday morning?”\newline
[Question]  Is S2 willing to present on Friday?\newline
[Choices] A Yes · B No · C Unsure · D Only afternoon\newline
[Audio context]  steady rain and mild line hiss \newline
[Ground truth] B \newline
[Simple model response] A \newline
[Reflection] \texttt{<REFLECT>}
..."Yeah, Friday… great" <PAUSE> check tone/intonation. </PAUSE> The phrase “Yeah, Friday… great” is produced with a downward F0 contour and reduced energy consistent with reluctance or irony, and a subsequent turn states explicit unavailability on Friday evening; integrating prosodic and semantic cues indicates S2 is not willing to present on Friday. \texttt{</REFLECT>}

\end{tcolorbox}

In sum, the Multi-speaker Tracking matters speaker attribution ("who"), the PAUSE Mechanism addresses the transition from rapid, text-centric processing to a more computationally intensive, multi-modal analysis, and Reflection focuses on the ground truth by deploying targeted evidence re-querying, contextual anchoring to disambiguate local hypotheses (as in ASR N-best lists), and integrating conflicting cross-modal data. The convergence of these mechanisms allows the model to emulate human-like cognitive robustness in complex, ambiguous, and multi-participant acoustic environments, distinguishing its performance from that of a passive, deterministic system.

\subsection{Self-correction Failure}
If the self-correction module fails, the model attempts to self-correct its initial text-biased assumption (Enthusiastic support) but misinterprets the conversational hedge and ignores the prosody, landing on the incorrect "safe" answer (Indifference).
\begin{tcolorbox}[
  breakable,
  enhanced,
  title=Bad Case D,
  colback=gray!10,
  colframe=gray!70!black,
  arc=1mm,
  boxrule=1pt,
  pad at break=1mm,
  left=1.5mm,
  right=1.5mm,
  top=1.5mm,
  bottom=2mm
]
\small%
Question: <audio>Which of the following best describes the speaker's most likely reaction upon seeing another picture of Ed Begley Jr. in an electric car based on the audio? (a) Mild amusement (b) Indifference (c) Exasperated sigh or groan (d) Enthusiastic support

Reply:\newline
\texttt{<THINK>}\newline
\texttt{<PLANNING>}Analyze the audio input to determine the speaker's emotional state and predict their reaction to seeing another picture of Ed Begley Jr. I need to extract the transcript, identify background sounds, and evaluate the speaker's vocal tone. \texttt{</PLANNING>}\newline
\texttt{<CAPTION>}
\texttt{<ENV>}Quiet indoor environment with no significant background noise.\texttt{</ENV>}
\texttt{<ASR>}I'm just saying, if I see one more picture of Ed Begley Jr. in that stupid electric car...\texttt{</ASR>}
\texttt{<SPEAKER>}Male voice, speaking at a normal volume, slightly drawn out pacing.\texttt{</SPEAKER>}
\texttt{</CAPTION>}\newline
\texttt{<REASONING>} The speaker is talking about Ed Begley Jr. and an electric car. Given that electric cars are generally associated with positive environmental efforts, the initial semantic assumption points toward (d) Enthusiastic support. \texttt{</REASONING>}\newline
\texttt{<SUMMARY>}The audio transcript involves a well-known environmentalist and an electric car.\texttt{</SUMMARY>}\newline
\texttt{</THINK>}
\end{tcolorbox}

Here is the reflection of this case.
\begin{tcolorbox}[
  breakable,
  enhanced,
  title=Bad Case D's reflection,
  colback=gray!10,
  colframe=gray!70!black,
  arc=1mm,
  boxrule=1pt,
  pad at break=1mm,
  left=1.5mm,
  right=1.5mm,
  top=1.5mm,
  bottom=2mm
]
\small%
\texttt{<REFLECT>} Wait, analyzing the transcript again, the speaker starts with "I'm just saying." This is a common conversational filler used to soften a statement or indicate a lack of strong commitment to the topic. Furthermore, the sentence trails off without a definitive aggressive conclusion. Therefore, enthusiastic support is incorrect. The passive nature of "I'm just saying" aligns best with a neutral, uninvested emotional state. I will pivot from enthusiastic support to indifference.\texttt{</REFLECT>}\newline
\texttt{<FINAL\_ANSWER>}Based on the use of passive conversational fillers like "I'm just saying," the speaker does not demonstrate strong emotional investment in the topic. The most likely reaction is (b) Indifference.\texttt{</FINAL\_ANSWER>}
\end{tcolorbox}

\section{Limitations of Simple ASR-Centric Text Reasoning}

\begin{figure*}[htp]
\centering
\includegraphics[width=0.9\linewidth]{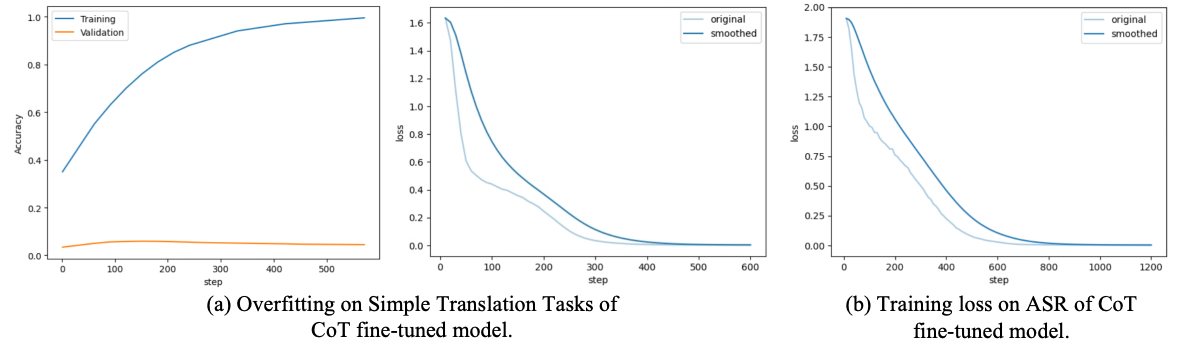}
\caption{The training dynamics of a chain-of-thought (CoT) fine-tuned model (Qwen2-Audio-7B-Instruct), indicating the model overfits to the training set in simple translation tasks. This suggests that CoT fine-tuning without additional regularization or more diverse data fails to yield robust generalization, particularly for tasks requiring broader reasoning beyond surface transcript matching.}
\label{fig:overfit}
\end{figure*}

Early approaches to audio reasoning typically relied on converting speech into text via automatic speech recognition (ASR) and then performing reasoning over the textual transcript. While effective to some extent, this paradigm inevitably discards information that is uniquely embedded in the audio signal itself. To probe the limitations of this pipeline, we first evaluated the ASR+text reasoning approach on benchmarks such as CoVoST2 and MMAU. In CoVoST2, model performance is largely determined by raw ASR accuracy, and we observed that “simple ASR” signals are quickly memorized without yielding robust generalization. 

Homophones and proper-name ambiguities necessitate long-range semantic modeling and external knowledge retrieval, while gendered pronouns in Chinese (e.g., “he/she”) lack reliable acoustic cues and thus require contextual inference for disambiguation. In particular, Paraformer’s frame-level alignment, coupled with strong language model priors, tends to induce a “nearest-neighbor copying” effect—yielding high accuracy on in-distribution transcripts but exhibiting pronounced failures under distributional shifts. Moreover, exposure to translation-oriented data (e.g., CoVoST2) can bias models such as Qwen-Audio to mistakenly trigger translation behavior, sometimes converting Chinese speech into other languages when acoustic cues are uncertain.

In Fig.~\ref{fig:appASRanalysis}(a), there is an improvement on base models if we asked them to answer questions with thinking in the format of \texttt{<THINK>...</THINK>} \texttt{<FINAL\_ANSWER>...</FINAL\_ANSWER>}. Therefore, we collected 2,050 samples from a subset of CoVoST2 (including 50 challenging cases reserved for the test set) and employed Kimi to generate CoT annotations.
Using this data, we fine-tuned Qwen2-Audio-7B-Instruct and evaluated them on the designated test set. However, the models exhibited severe overfitting (see Fig.~\ref{fig:overfit}(b)) after only a single epoch of training: while the outputs consistently followed the required \texttt{<THINK>...</THINK>} \texttt{<FINAL\_ANSWER>...</FINAL\_ANSWER>} format and the training loss rapidly approached zero, the test accuracy dropped below 5\%. This observation indicates that the gradients primarily optimized for surface-level grapheme mapping and fixed output formatting, without fostering genuine cross-sentence reasoning, coreference resolution, or knowledge-grounded inference.

Consequently, these observations indicate that the “Thinking” component of chain-of-thought supervision should be allocated primarily to more challenging audio understanding tasks, such as multi-speaker dialogues and noisy environments, where reasoning signals drive the model to overcome semantic ambiguities and enforce knowledge-aware interpretations, rather than merely replicating templates on simple ASR tasks.

\section{The Use of Large Language Models (LLMs)}
In order to reduce typos during the writing process and to optimize complex sentence structures so that the article becomes simpler and easier to read, we use mainstream large language models to refine certain paragraphs. For example, we use prompts such as “Help me correct the typos and grammatical errors in the above text, and streamline the logic to make it clear and easy to understand.”

\end{document}